\documentclass[showpacs,preprintnumbers,amsmath,amssymb]{revtex4}

\usepackage{graphicx}
\usepackage{dcolumn}
\usepackage{bm}

\begin{document}

\preprint{APS/123-QED}

\title{On a multi-scale approach to analyze the joint statistics of longitudinal and transverse increments experimentally in small scale turbulence}

\author{M. Siefert}
\author{J. Peinke}%
 \email{peinke@uni-oldenburg.de}
  \homepage{http://www.uni-oldenburg.de/hydro}
\affiliation{%
Faculty V, Institute of Physics, Carl-von-Ossietzky University of Oldenburg
}%


\date{\today}

\begin{abstract}
We analyze the relationship of longitudinal and transverse increment statistics measured in isotropic small-scale turbulence. This is done by means of the theory of Markov processes leading to a phenomenological Fokker-Planck equation for the two increments from which a generalized K\'arm\'an equation is derived. We discuss in detail the analysis and show that the estimated equation can describe the statistics of the turbulent cascade. A remarkably result is that the main differences between longitudinal and transverse increments can be explained by a simple rescaling-symmetry, namely the cascade speed of the transverse increments is 1.5 times faster than that of the longitudinal increments. Small differences can be found in the skewness and in a higher order intermittency term. The rescaling symmetry is compatible with the Kolmogorov constants and the K\'arm\'an equation and give new insight into the use of extended self similarity (ESS) for transverse increments. Based on the results we propose an extended self similarity for the transverse increments (ESST).
\end{abstract}

\pacs{05.10.Gg, 47.27.-i, 47.27.Eq, 47.27.Gs, 47.27.Vf}
\maketitle
\section{Introduction}

Small scale turbulence is not yet full understood. A complete theory based on the Navier-Stokes equation has not been achieved yet, thus our present understanding relies for the most part on phenomenological and experimental approaches. It is assumed that turbulence forms an {\em universal} state which exhibits stationarity, isotropic and homogeneity in a statistical sense \cite{Monin75,frisch95}. In general, turbulence are driven on large scale, i.e. energy is injected into large scale motions and is dissipated on small scales resulting in a net flux of energy from large to small scales \cite{richardson20}. The energy flux results from the inherent instability and the subsequent breakup of vortices into smaller ones.

For local isotropic turbulence the main challenge is to understand spatial correlation. Usually the turbulent field $\bf{U}(x,t)$ is characterized by increments 
$u_r={\bf e}\cdot\left[ {\bf U}({\bf x}+{\bf r},t)-{\bf U}({\bf x},t)\right]$,
where ${\bf e}$ denotes a unit vector in a certain direction, ${\bf x}$ denotes a reference point and $\mathbf{r}$ a displacement vector. The increments are taken as stochastic variables in dependence of the scale variable $r=|\mathbf{r}|$ and by varying $r$ correlation on different scales can be studied with the increments. In the following, we assign $u_r$ to longitudinal increments, that means ${\bf e}$ is parallel to ${\bf r}$ and $v_r$ to transversal increments, for which ${\bf e}$ is orthogonal to ${\bf r}$. For specific length $r_i$ we write $u_i$ and $v_i$ in a short way. The central challenge in turbulence is to explain the statistics of exceptional frequent occurrence of large velocity increments on small scales $r$, which can not be understood with normal statistics. This is the so called intermittency problem.

The work of Kolmogorov \cite{kolmogorov41a,kolmogorov41c} is still the foundation of the small-scale turbulence theory.
The starting hypothesis is that the possible symmetries of the Navier-Stokes equation are recovered in a statistical sense for high Reynolds numbers. These symmetries are homogeneity, i.e. the statistics of the increments is independent of the reference point ${\bf x}$, isotropy, i.e. the statistics does not change under rotation of the frame of reference, and stationarity, i.e. the statistics is not time dependent. A further hypothesis is scale invariance, i.e. loosely spoken the structures of different sizes looks similar. Kolmogorov has considered the statistics in terms of structure functions (moments of the velocity increments), for which scale-invariance reads as $\langle u_{r}^n\rangle=\left( r/r_1\right)^{\xi_n}\langle u_{1}^n\rangle$ implying $\langle  u_{r}^n\rangle\propto r^{\xi_n}$ with two different distances $r$ and $r_1$.

Using this hypothesis, Kolmogorov has furthermore assumed that for high Reynolds number the statistics of velocity increments depends only on the energy dissipation $\epsilon$ and the scale $r$ and ended up by using dimension arguments with $\langle u_{r}^n \rangle=C_{n}\epsilon^{n/3}r^{n/3}$ for $\eta\ll r\ll L$. The constants $C_n$ denote the Kolmogorov constants, $\eta$ is the scale where the dissipation takes place and $L$, the integral length scale, is the largest scale of the flow. However there are correction due to the fluctuating dissipation energy leading to deviations from the exponent $\xi_n=n/3$. This typical property of turbulence is an other aspect of the above mentioned intermittency. According to the refined self similarity hypothesis (RSH) of Kolmogorov, which takes into account the fluctuating energy dissipation $\left<\epsilon_r\right>$ averaged over a volume $V$ with the extension $r$ ($\epsilon_r=\int_V \epsilon dx$), the scale-invariance is given by
\begin{equation}\label{scaling}
\langle u_{r}^n \rangle=c_{n}\langle\epsilon_r^{n/3}\rangle r^{n/3} \propto r^{\xi_n}.
\end{equation} 
The first model for the exponents $\xi_n$ is Kolmogorov's log-normal model, which results in $\xi_n=n/3-\mu n(n-3)/18$ \cite{kolmogorov62}. If instead of the structure functions the probability density functions (pdf) $p(u_r)$ are considert, this nontrivial scaling behavior correspond to a change of the form of the pdfs with $r$. To explain intermittency, i.e. the non-trivial scaling behaviour, is still one of the prominent challenges in turbulence.

At first, the examinations were only concentrated on the longitudinal increments. There was hope that the exponent $\xi_n$ can describe the self-similarity of every quantity of the velocity field. But recently, a lot of afford was undertaken to consider also the transverse increments and they seem to have essential different properties \cite{herweijer95b,lvov96f,camussi96b,kahalerras96,pearson97,noullez97,grossmann97a,grossmann97b,camussi97,chen97a,boratav97c,boratav97b,boratav97a,antonia97c,chen97a,sreenivasan97,dhruva97,chen97b,he98,kahalerras98,pedrizzetti99,vandewater99,antonia99,nelkin99,antonia00a,zhou00,malecot00,nelkin00,romano01}. This implies that the models as well as the considerations concerning only the longitudinal increments are too specific and a better understanding of the turbulence has to include the transverse increments.

Longitudinal and transverse increments belong to different geometric / kinematic structures in the flow. The longitudinal increments can be associated with strain-like structures, the transverse increments with vorticity-like structures. Thus it is natural to modify RSH of equation (\ref{scaling}) and to use local averaged enstrophy (squared vorticity) rather than local averaged dissipation for the transverse increments (refined similarity hypothesis for transverse increments, RSHT, see  \cite{chen97a,boratav97c}), 
\begin{equation}
\langle v_{r}^n \rangle=C_{\textrm{t},n}\langle\Omega_r^{n/3}\rangle r^{n/3} \propto r^{\xi_{t,n}}.
\end{equation}
If intermittency results from the fluctuating dissipation and enstrophy, then the deviations from the Kolmogorov 41 theory \cite{kolmogorov41c} can be investigated by the scaling of $\langle\epsilon_r^{n/3}\rangle$ and $\langle\Omega_r^{n/3}\rangle$, respectively. For infinite high Reynolds number the scaling of averaged dissipation and enstrophy should be the same \cite{nelkin99}, but in many experiments and simulation one observe differences in both exponents for finite Reynolds number \cite{herweijer95b,pearson97,boratav97c,grossmann97a,camussi97,boratav97a,dhruva97,zhou00,nelkin00,romano01,Tsinober01}. At least four arguments are used to explain this observations: 1) anisotropy, 2) effect of Reynolds number, 3) influence of boundary conditions, 4) intermittency. Anisotropy typically exists in every flow on large scales and it can influence the exponent stronger than intermittency itself \cite{zhou00,romano01}. 
 Decreasing the Reynolds number effects the scaling exponents in such a way that the differences between them increase \cite{dhruva97,zhou00,antonia00a,romano01}.

The structure functions $\langle u_r^n\rangle$ and $\langle v_r^n\rangle$ as well as the scaling properties involved can not define unambiguously the turbulent field; for instance flows with different structures may have the same scaling properties \cite{Tsinober01}. The same is true for the probability distributions $p(u_r)$ and $p(v_r)$, which are in essential equivalent to the structure functions. The reason is that these quantities are just two-point statistics. Definitive more general and detailed are the multi-point (or multi-scale) distributions $p(u_1,v_1,r_1;\dots;u_n,v_n,r_n)$ (or multi-scale structure functions $\langle\prod_i u^{m_i}_i\prod_j v^{m_{j}}_j\rangle$) for different scales $r_i$. These probabilities also consider the joint statistics of longitudinal and transverse increments and thus enables also to describe the interaction between them. Furthermore they describe the simultaneous occurrence of increments $u_i,\,v_i$ on several length scales $r_i$.

In this paper, we focus on an approach to characterize multi-scale statistics. It has been shown that it is possible to get access to the joint probability distribution $p(u_1,r_1;u_2,r_2;\dots;u_n,r_n)$ via a Fokker-Planck equation estimated directly from measured data \cite{friedrich97a,friedrich97b,marcq01}. This has attracted interest and was applied to different problems of the complexity of turbulence like energy dissipation \cite{naert97,marcq98,hosokawa02}, universality of turbulence \cite{renner02}, the theoretical derivation of the Fokker-Planck-equation from the Navier-Stokes equation in the limit of high Reynolds number \cite{davoudi99} and the analysis of stochastic time series \cite{siegert98,friedrich00a,siefert03}. The characteristic of this method is that it is based on pure data analysis, i.e. it is a parameter-free method, and that the few underlying assumptions are verifiable. Thus no special model-ideas are interwoven with the procedure.

In this article we extend this approach to analyze the joint statistics of longitudinal and transverse increments. A first result concerning the different cascade speed of longitudinal and transverse cascade has been published in a preceding letter \cite{siefert04}. The aim of this article is to present the used method and its extension to longitudinal and transverse increments in detail and to discuss the similarities and differences of longitudinal and transverse increments. 
We start with a short description of the concepts of Markov processes and define the notation. In section \ref{sec:experiment} we describe the experiment and characterize the data with two-point statistics (i.e. with one scale) in section \ref{sec:twopoint}. Next, we analyze longitudinal and transverse increments separately by means of Markov-processes with respect to multi-points or multi-scale statistics. Then we go over to a combined analysis of both increments and discuss a new symmetry which connect both increments. An interpretation in section \ref{sec:discussion} and a conclusion will finish the paper.

\section{The mathematics of Markov processes}\label{theory}

The basis of our argumentation and of the used analysis is the theory of Markov and diffusion processes. Therefore we will give a concise description of them, which also serve as a guideline for the analysis of the turbulent signal. The foundations are known since Kolmogorov \cite{kolmogorov31}, but a detailed presentation can also be found in standard textbooks such as \cite{Risken,Gardiner,haenggi82}.

We restrict ourself to the case of a two dimensional stochastic variable, denoted by the stochastic vector
\begin{equation}
{\mathbf u_r}:=\left(\begin{array}{c}
u_r\\v_r
\end{array}\right).\label{ustate}
\end{equation}
Usually, a stochastic process is formulated in the time $t$ but one can also imagine different independent scalars; for our purpose to characterize the turbulent cascade the scale $r$ is the independent variable.

The stochastic process underlying the evolution of ${\mathbf  u}_r$
in the scale $r$ is Markovian, if the conditional pdf $p\left( \, 
{\mathbf  u}_{1},r_1 \, | \, {\mathbf  u}_{2},r_2;
\dots;{\mathbf  u}_{n},r_n, \, \right)$ with $r_{1}\leq r_{2} \leq \dots  \leq
r_{n}$ fulfills the relation
\begin{equation}
	p\left( \, {\mathbf  u}_1,r_1 \, | \, {\mathbf  u}_2,r_2;
	{\mathbf  u}_3,r_3; \dots ; {\mathbf  u}_n,r_n \, ) = p( \,
	{\mathbf  u}_1 ,r_1\, | \, {\mathbf  u}_2 ,r_2\, \right) ,
	\label{MarkovCond}
\end{equation}
where $p\left( \, {\mathbf  u}_1,r_1 \, | \, {\mathbf  u}_2,r_2,{\mathbf
u}_3,r_3; \dots ; {\mathbf  u}_n,r_n \, \right)$ denotes the
probability for finding certain values for ${\mathbf u}_1$ at some scale
$r_{1}$, provided that the values of ${\mathbf  u}_i$ at the larger
scales $r_i>r_1$, $i>1$, are known. The condition
(\ref{MarkovCond}) states that the increment distribution of $u_1$ on $r_1$ only depends on the increment value on one larger scale $r_2$ and that further scales do not give more information. Markov properties imply a remarkable property. The knowledge of the
conditional pdf $p\left( \, {\mathbf u}_r ,r\, | \, {\mathbf
u}_0,r_0 \, \right) $ (with
$r \leq r_{0}$) is sufficient to determine any $n$--point pdf:
\begin{eqnarray}
	p\left( \, {\mathbf u}_1,r_1,{\mathbf u}_2,r_2,\dots, {\mathbf
	u}_n ,r_n\, \right) & = & p \left( \, {\mathbf u}_1,r_1 \,
	| \, {\mathbf u}_2, r_2\, \right) \times p\left( \, {\mathbf
	u}_2,r_2 \, | \, {\mathbf u}_3,r_3 \, \right) \times ... 
	\nonumber \\
	& & \times p\left( \, {\mathbf u}_{n-1},r_{n-1} \, | \, {\mathbf
	u}_{n},r_n \, \right) \times p\left( \, {\mathbf u}_{n},r_n \,
	\right),
	\label{chain}
\end{eqnarray}
i.e. the single conditioned pdfs contain the entire information about the stochastic process

For Markov processes the evolution
of the conditional pdf in the scale $r$ can be described by the
Kramers-Moyal expansion, a partial differential equation for $p\left(
\, {\mathbf u}_r ,r\, | \, {\mathbf u}_0,r_0 \, \right)$ in the
variables ${\mathbf u}_r$ and $r$.  According to Pawula's theorem, this
expansion truncates after the second term if the fourth order
expansion coefficient vanishes \cite{Risken}.  In this case, the Kramers-Moyal
expansion reduces to the Fokker--Planck equation (or Kolmogorov equation)
\begin{eqnarray}
	- r \frac{\partial}{\partial r} p( \, {\mathbf  u},r \, | \,
	{\mathbf  u}_0,r_{0} \, ) 
	& = & -  \sum\limits_{i=1}^{2}
	\frac{\partial}{\partial u_{i}} \left( \;
	D^{(1)}_{i}({\mathbf  u},r) \; p(\, {\mathbf  u},r \, |\,
	{\mathbf  u}_0,r_{0}) \; \right) 
	\label{FoplaCond}\\
	&& +  \sum\limits_{i,j=1}^{2} \frac{\partial^2}{\partial
	u_{i} \partial u_{j}} \left( \; D^{(2)}_{ij}
	({\mathbf  u},r) \; p( \, {\mathbf  u},r \, | \, {\mathbf
	u}_0,r_{0} \, ) \; \right) .
	\nonumber
\end{eqnarray}
Note that we have multiplied both sides of the usually used Fokker-Planck equation with $r$; on the right hand side we have incorporated this factor in the definition of the Kramers-Moyal coefficients (\ref{MkDef}). The Fokker-Planck equation describes not only the evolution of the conditional probability $p( \, {\mathbf  u}_{r},r \, | \, {\mathbf u}_{r_{0}},r_{0} \, )$ but also the evolution of the pdf $p({\mathbf u}_r,r)$, as can bee seen by integrating (\ref{FoplaCond}) over $u_0$.

The drift vector ${\mathbf D}^{(1)}$ and the diffusion matrix ${\mathbf D}^{(2)}$ of (\ref{FoplaCond}) are
defined via the limit
\begin{eqnarray}
    D^{(1)}_{i}({\mathbf  u},r)  & = & \lim_{\Delta r \rightarrow 0} \,
    M^{(1)}_{i}({\mathbf  u},r, \Delta r) , \nonumber \\
    D^{(2)}_{ij}({\mathbf  u},r)  & = & \lim_{\Delta r \rightarrow 0} \,
    M^{(2)}_{ij}({\mathbf  u},r,\Delta r) , \label{DkDef}
\end{eqnarray}    
where the coefficients ${\mathbf M}^{(k)}$ are given by
\begin{eqnarray}
	M^{(1)}_{i}({\mathbf u},r,\Delta r) & = & \frac{r}{\Delta r}
	\left< \; \left.  \left( \, u_{i}'(r-\Delta r) - u_{i}(r) \,
	\right) \, \right| \, {\mathbf u},r \; \right> \; , \nonumber
	\\
	M^{(2)}_{ij}({\mathbf  u},r,\Delta r) & = & \frac{r}{2 \Delta
	r} \left< \; \left( \, u_{i}'(r-\Delta r) - u_{i}(r)
	\right) \right.  \, \times 
	\nonumber \\
	&& \left.  \left.  \times \, \left( \, u_{j}'(r-\Delta r) -
	u_{j}(r) \, \right) \right| {\mathbf  u},r \; \right> . 
	\label{MkDef}
\end{eqnarray}
The coefficients ${\mathbf M}^{(k)}$ are conditional
expectation values $\left<\cdot|{\mathbf u}\right>$ on the veloctiy increment ${\mathbf u}$ and can easily be determined from experimental data.  One can find estimates for the
${\mathbf D}^{(k)}$ by extrapolating the measured conditional moments ${\mathbf M}^{(k)}$ in dependence of $\Delta r$
 towards $\Delta r = 0$ \cite{renner01a}. An other possibility would be to approximate the drift and diffusion coefficients by the coefficients  ${\mathbf M}^{(k)}$ for one finite $\Delta r$. The corrections of higher order in $\Delta r$ can then be taken into consideration by correction terms. The approved coefficients can be used to recalculate the corrections, and if one perform this procedure recursively, the result converge towards the drift and diffusion coefficients \cite{ragwitz01}. Both approaches give similar results, but the approximation of \cite{ragwitz01} has some pitfalls, if it does not examined the limes, which is an essential point of this analysis, see also \cite{friedrich02}.

A solution $p({\mathbf u},l)$ of the Fokker-Planck equation can be derived from the Chapman-Kolmogorov equation $p({\mathbf u},l)=\int p({\mathbf u},l|{\mathbf u}_0,l_0)p({\mathbf u}_0,l_0)d{\mathbf u}_0$. Here we use the logarithmic scale $l:=\ln(L/r)$ to transform the Fokker-Planck equation into the usual form $\partial p/\partial l=\dots$ instead of $-r\partial p/\partial r$=\dots. For sufficient small $\Delta l\equiv l-l_0$, such that $D^{(i)}$ are constant in $\Delta l$, the conditioned probability can be approximated by 
\begin{eqnarray}\label{smalldr}
\lefteqn{p({\bf u},l+\Delta l|{\bf u'},l)\approx\frac{1}{4\pi\sqrt{\det{\bf D^{(2)}({\bf u'},l)}\Delta l}}\times}\\&&\times\exp\left(-\frac{1}{4\Delta l}\left[\left({{\bf D}^{(2)}({\bf u'})}\right)^{-1}\right]_{jk}\left(u_j-u'_j-D^{(1)}_j\Delta l\right)\left(u_k-u'_k-D^{(1)}_k\Delta l\right)\right).\nonumber
\end{eqnarray}
For larger distances $l-l_0$, a solution can be obtained by reiterating the Chapman-Kolmogrov equation with this approximation:
\begin{equation}
p(\mathbf{u},l)=\int d\mathbf{u}_{N-1}\dots \int d\mathbf{u}_0\;p(\mathbf{u},r|\mathbf{u}_{N-1},l_{N-1})\cdots p(\mathbf{u}_1,l_1|\mathbf{u}_0,l_0)p(\mathbf{u}_0,l_0).\label{pathint}
\end{equation}
Below we use this path-integral approximation to solve the Fokker-Planck equation numerically. 

The error analysis for a diffusion process in two variables is more complicated than for one dimension. We perform the error analysis according to \cite{siegert01}; here we just sketch the procedure and give the resulting equations for the error estimations. An error analysis comprises  a quantity to estimate and a stochastic quantity which leads to an uncertainty. The two quantitates to estimate are the drift and diffusion coefficients and it is assumed that the stochastic quantity is given solely due to the stochastic nature of the process. Because the randomness is determined by $D^{(2)}$, the error of $D^{(1)}$ and $D^{(2)}$ can be expressed by $D^{(2)}$ itself. The following results are valid for small $\Delta r$. The error of the drift coefficient is given by
\begin{equation}
\Delta D_i^{(1)}=\frac{1}{\sqrt{N\Delta l}}\sqrt{D_{ii}^{(2)}},\label{errorD1}
\end{equation}
where $N$ is the number of samples used to estimate $D^{(1)}$. To calculate the error for the diffusion matrix, one has to transform the diffusion coefficient by a suitable orthogonal transformation matrix $\mathbf{B}$ in the diagonal form. In the diagonal system one has
\begin{equation}
\widetilde{\mathbf{D}}^{(1)}=\mathbf{B}\mathbf{D}^{(1)},\;\;\widetilde{\mathbf{D}}^{(2)}=\mathbf{B}\mathbf{D}^{(2)}\mathbf{B}^T.
\end{equation}
 The error of $\widetilde{D}^{(2)}_{ii}$ is then given by
 \begin{equation}\label{errorD2}
\Delta\widetilde{D}^{(2)}_{ii}=\widetilde{D}^{(2)}_{ii}\sigma(m=\widetilde{D}^{(1)}_{i}\sqrt{r}/\sqrt{\widetilde{D}^{(2)}_{ii}},N),
\end{equation}
which leads to the error of the initial diffusion matrix by the inverse transformation $\Delta\mathbf{D}^{(2)}=\mathbf{B}^T\Delta\widetilde{\mathbf{D}}^{(2)}\mathbf{B}$. The function $\sigma(m,N)$ is defined by 
\begin{equation}
\int\limits^{\left<\zeta\right>+\sigma}_{\left<\zeta\right>-\sigma}p(\widetilde{\zeta})d\widetilde{\zeta}=0.68,
\end{equation}
where 
\begin{equation}\widetilde{\zeta}=\frac 1N\sum_{i=1}^{N}\zeta_i
\end{equation}
 is an estimator for the average value of  a $\chi^2$-distributed stochastic variable $\zeta=(m+\Gamma)^2$ with a normal distributed stochastic variable $\Gamma$ obeying $\left<\Gamma\right>=0$ and $\left<\Gamma^2\right>=1$. Thus $\sigma(m,N)$ defines the 32\%-confidence interval of the estimator $\widetilde{\zeta}$. 
 
In the above presentation the central quantity was the pdf but very often one is interested in the structure functions $\langle u^mv^n\rangle=\int u^mv^n p(u,v,r)dudv$. From the Fokker-Planck equation a hierarchical equation for the structure functions can be derived by using $\langle u^mv^nf(u,v)\rangle=\int u^mv^n f(u,v)p(u,v,r)dudv$:
\begin{eqnarray}\label{StrukturGl}
\lefteqn{-\frac{\partial}{\partial r}\langle u^mv^n\rangle =}\\
&+&m \langle u^{m-1}v^n D^{(1)}_1(u,v,r) \rangle+n\langle u^mv^{n-1}D^{(1)}_2(u,v,r) \rangle\nonumber\\&+&\frac{m(m-1)}{2}\langle u^{m-2}v^n D^{(2)}_{11}(u,v,r)\rangle\nonumber\\&+&\frac{n(n-1)}{2}\langle u^{m}v^{n-2} D^{(2)}_{22}(u,v,r)\rangle \nonumber\\ &+&mn\langle u^{m-1}v^{n-1}D^{(2)}_{12}(u,v,r)\rangle.\nonumber
\end{eqnarray}
This equation allows a direct comparison of the Fokker-Planck equation with structure functions.

 \section{The experimental setup}\label{sec:experiment}

For the subsequent analysis we use two data sets measured in the central region of a wake behind a cylinder. The Reynolds numbers are $R_\lambda=180$ and $R_\lambda=550$, respectively. If nothing else is said, the data set with $R_\lambda=180$ is used. The high Reynolds number data set is used for comparison.

For the first data set the local velocity is measured in a wake 60 diameters behind a cylinder with cross section $D=20\;$mm. The Reynolds number is 13200 with a Taylor based Reynolds number of $R_\lambda=$180. We have measured the velocity component $U$ in the mean flow direction, the $V$ component orthogonal to the cylinder axis with an X-wire (Dantec's frame 90N10 with Dantec's 55P51 X-wire). We collect $1.25\cdot10^8$ samples with a sampling frequency of 25 kHz using a 16bit A/D converter; high frequency electronic noise was suppressed with a low-pass filter to prevent aliasing.  We use Taylor's hypothesis of frozen turbulence to convert time lags into spatial displacements.  With the sampling frequency of 25 kHz and a mean velocity of 9.1 m/s, the spatial resolution of the measurement is not better than 0.36 mm. For the integral length scale $L=\int_{0}^{r_{0}}R(r)dr$, where $r_{0}$ represent the first zero-crossing of the autocorrelation $R(r)$ function, we obtain a value of $L=$137$\;$mm for the stream-wise component $U$ and $L_t=$125$\;$mm for the component $V$. Taylor's microscale $\lambda$ is calculated using the method proposed by \cite{aronson93}. For isotropic turbulence, $\lambda$ can be written as
\begin{equation}
\lambda^{2}=\frac{\sigma}{\left<(\partial U/\partial x)^{2}\right>}=\lim_{r\to 0}\frac{\sigma r^{2}}{\left<  u^2_r\right>},
\end{equation}
where $\sigma$ is the standard deviation of the turbulent fluctuations.
The limit has been calculated by fitting a second-order polynomial to the data, resulting in 
 $\lambda=$4.8$\;$mm for the stream-wise component and $\lambda_t=$3.0$\;$mm for the $V$-component. The dissipation scale $\eta$ can not be resolved, because it is smaller than the length of the hot-wire and smaller than the spatial resolution calculated with the finite sampling frequency and the Taylor-hypothesis.

The high Reynolds number flow is measured 40 diameter behind a cylinder with a cross section of 5 cm. The Reynolds number is Re$=84600$, the Taylor based Reynolds number is $R_\lambda=550$. The $U$-component is  measured in direction of the mean flow, the $V$-component in direction of the cylinder axis. $1.25\cdot10^7$ data points are measured with a frequency of 200 kHz. The facilities are the same as in  the first measurement. Typical length are $L=15.5\;$cm and $\lambda=3.87\;$mm

 \section{Experimental results}\label{sec:twopoint}

\subsection{Two point statistics}

First of all, we characterize our data by two-point statistics (structure functions, pdf of increments etc.). In this way, we have a reference to results presented frequently in the literature and we get a first impression of the differences between longitudinal and transverse increments.
 
\begin{figure}[tbp]
\begin{center}
\includegraphics[width=3.5in]{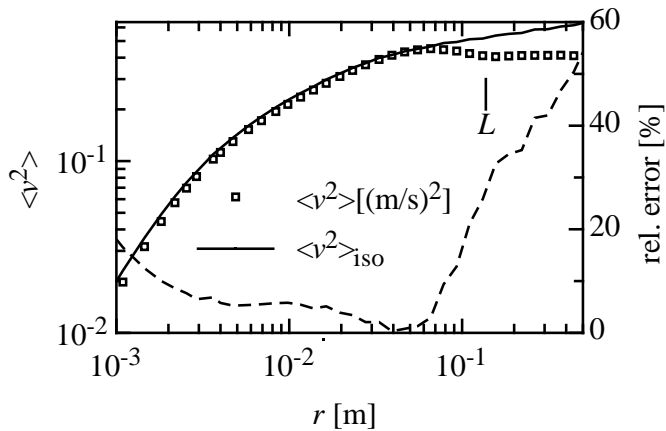}
\end{center}
\caption{Second order transverse structure function estimated on the one hand directly from data (squares) and on the other hand from the longitudinal structure function via the K\'arm\'an equation (straight line); i.e. the left and right hand side of equation (\ref{karman1}) are plotted. The relative deviations are plotted by a dashed line. For smaller scales $r<L$ the K\'arm\'an equation is well fulfilled within 5\% and is isotropic in this sense. On larger scales the data are clear anisotropic. \label{fig:karman} }
\end{figure}

A central pre-assumption for many considerations of small scale turbulence is isotropy. To study isotropy we use the K\'arm\'an equation (isotropy relation) because it is the simplest exact relation for structure functions which rely on isotropy. The K\'arm\'an equation connects the second order longitudinal and transversal structure functions:
 \begin{eqnarray}\label{karman1}
\left<v_r^2\right>=\left<u_r^2\right>+\frac{r}{2}\frac{\partial}{\partial r}\left<u_r^2\right>\;\;\;\textrm{first K\'arm\'an equation}.
\end{eqnarray}
Fig. \ref{fig:karman} shows the left and right hand side of the K\'arm\'an equation, i.e. the second transverse structure function and the hypothetic transverse structure function for isotropic flows calculated from the longitudinal one. Taking the validity of the K\'arm\'an equation as indication of isotropy, we see that for small distances $r<L$ the isotropic relation is well fulfilled, the relative error is within 5\%. For large distances $r>L$ isotropy is violated (large scale anisotropy), the ratio $\langle v_r^2\rangle/\langle u_r^2\rangle$ is approximately 0.66, which is a typical value in the literature.

\begin{figure}[tbp]
\begin{center}
\includegraphics[width=2.5in]{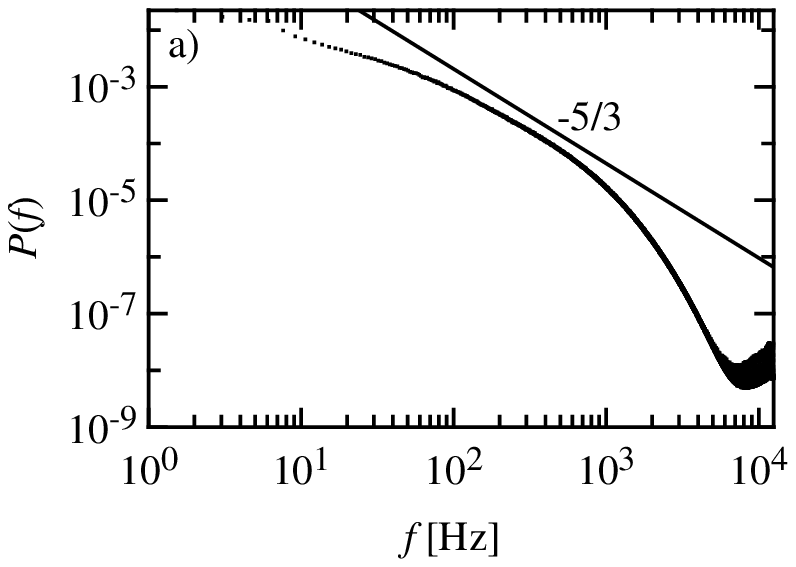}
\includegraphics[width=2.5in]{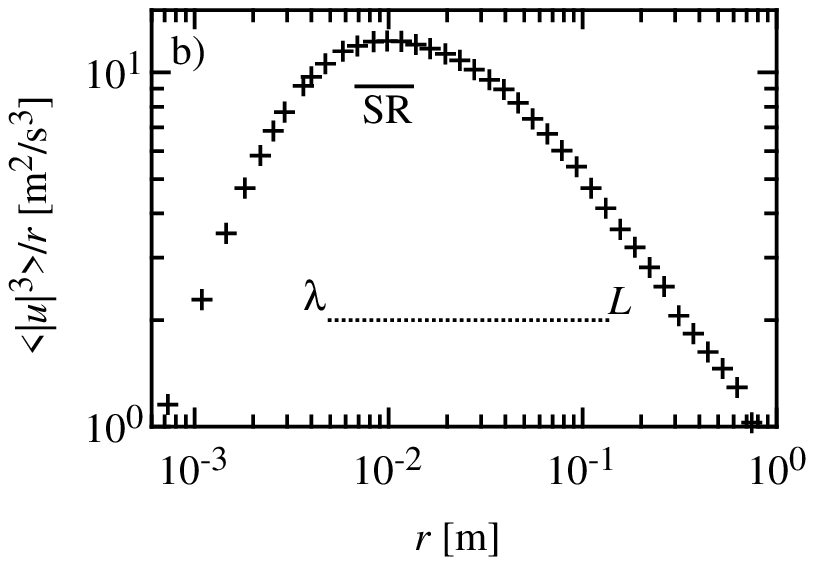}
\end{center}
\caption{a) Energy spectrum (dots) with a -5/3--power law and b) compensated plot of the third order structure function. The plateau of $\left<|u_r|^3\right>/r$ defines the scaling range SR. (Data for R$_\lambda$=180.)\label{fig:scaling}}
\end{figure}

Next, we analyze the data with respect to scaling properties. Fig. \ref{fig:scaling}a) shows the energy spectrum with Kolmogorov's -5/3--law for comparison. In Fig. \ref{fig:scaling}b) the third order structure function is plotted in a compensated representation, i.e. $\langle |u_r^{3}|\rangle/r$ is plotted against $r$ to estimate the scaling range. The maximum lies at $10^{-2}\;$m, which defines according to Kolmogorov's 4/5-law the position and the width of the scaling range.

\begin{figure}[tbp]
\begin{center}
\includegraphics[width=1.8in]{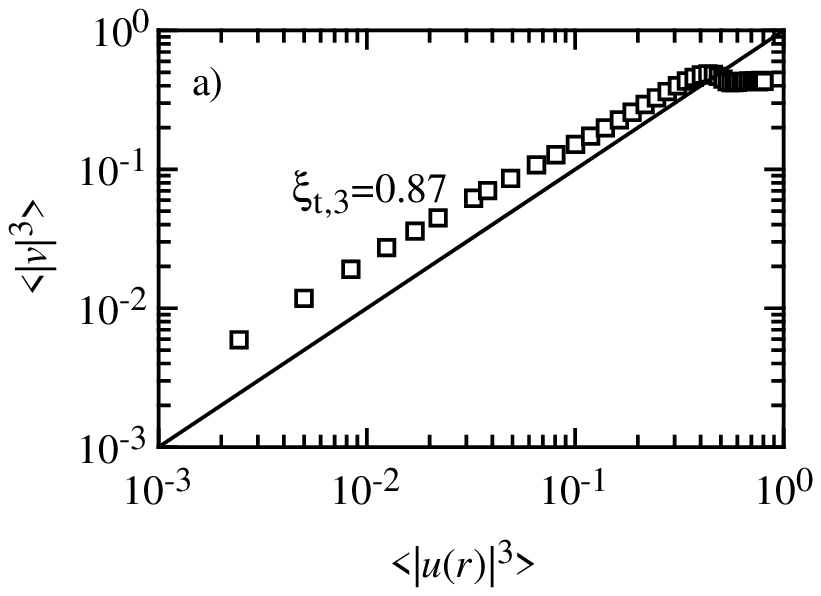}
\includegraphics[width=1.8in]{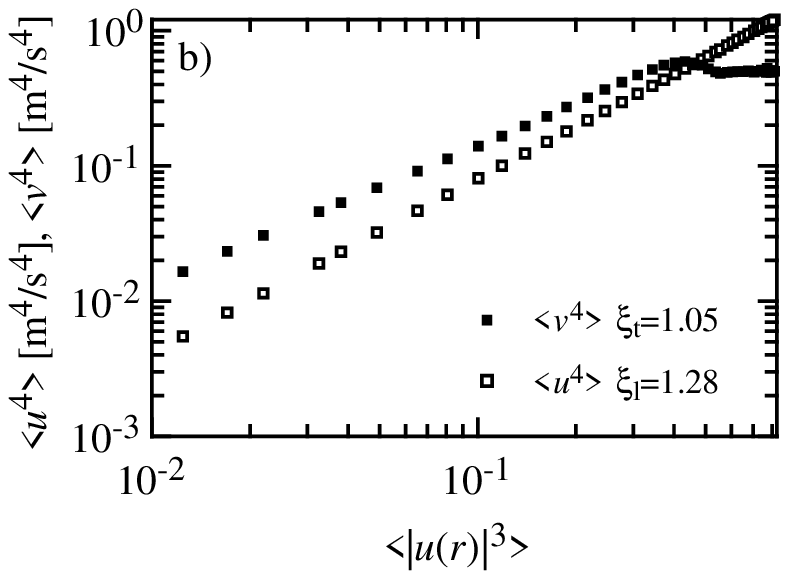}
\includegraphics[width=1.8in]{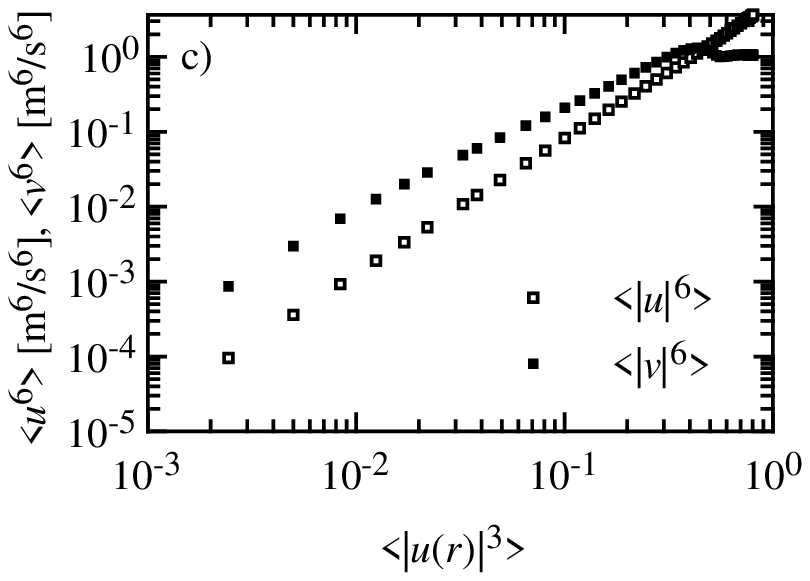}
\end{center}
\caption{The ESS representation of the longitudinal and transverse structure functions. a) The transverse third order structure function is plotted versus the longitudinal third order structure function. The exponent is $\xi^{t}_3=0.87$. The line has an exponent of 1, which is expected from Kolmogorov's 4/5-law for the third order longitudinal structure function.  b) The fourth order structure functions and c) the sixth order structure functions. Between the longitudinal and transverse structure functions clear differences can be seen. \label{fig:ESS} }
\end{figure}

Because the scaling range is narrow, we use the extended self similarity (ESS) \cite{benzi93b,benzi95}, see also appendix \ref{appendixa},
\begin{equation}
\left<|u_r|^n\right>\propto\left<|u_r|^3\right>^{\xi_n^l}\label{ESSa1}
\end{equation}
\begin{equation}
\langle |v_r|^n\rangle\propto \langle|u_r|^3\rangle^{\xi_n^t}\label{ESSb1}
\end{equation}
to estimate the scaling exponents of the structure functions. In Fig. \ref{fig:ESS} we show the application of ESS to the third, fourth and sixth order structure functions. The third order structure function is of especially interest, because it serve as the reference in ESS. In all three figures it can be seen that the transverse exponent is smaller than the longitudinal one, $\xi_n^{t}< \xi_n^{l}$. This result is well accepted \cite{pearson01,dhruva97,antonia97a,zhou00,chen97a}. As a consequence, one scaling group is not enough to characterize the turbulent flow and the statistics is more complex than previous thought.

\begin{figure}[tbp]
\begin{center}
\includegraphics[width=3.5in]{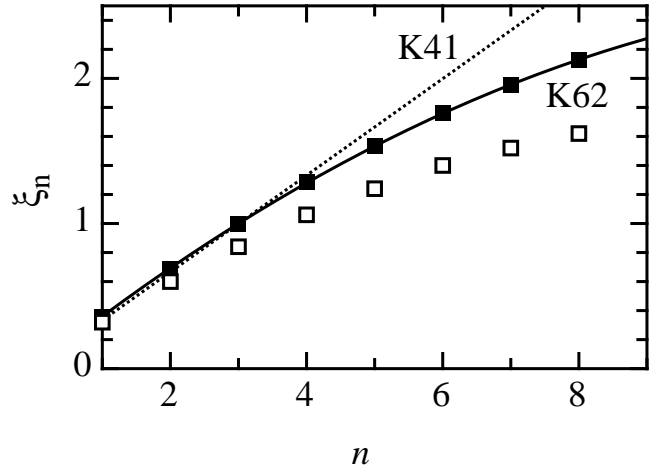}
\end{center}
\caption{\label{fig:ESSb} The scaling exponents estimated with ESS. Bold squares: the scaling exponents $\xi_n^l$ up to order 8. The dotted line is the forecast of K41, the straight line includes the intermittency corrections of K62 with the fitted intermittency parameter $\mu=0.24$. Open squares: transverse exponents.}
\end{figure}

To make this more clear, Fig. \ref{fig:ESSb} shows the exponents $\xi_n^l$ and $\xi_n^t$ up to order 8. For the sixth order structure function we get $\xi_6^l= 1.76\;\pm 0.04$. Fitting Kolmogorov's log-normal model to the values of $\xi_n^l$ yields the intermittency parameter $\mu=$0.24. Both is in good accordance with the experimentally expected values, see for example \cite{arneodo96}. The transverse exponents are clearly smaller then the longitudinal exponents for $n>3$, i.e. the transverse structures are significantly more intermittent, see Fig. \ref{fig:ESSb}. This heavily discussed point will be discussed in detail at the end of this article.

\begin{figure}[tbp]
\begin{center}
\includegraphics[width=3.5in]{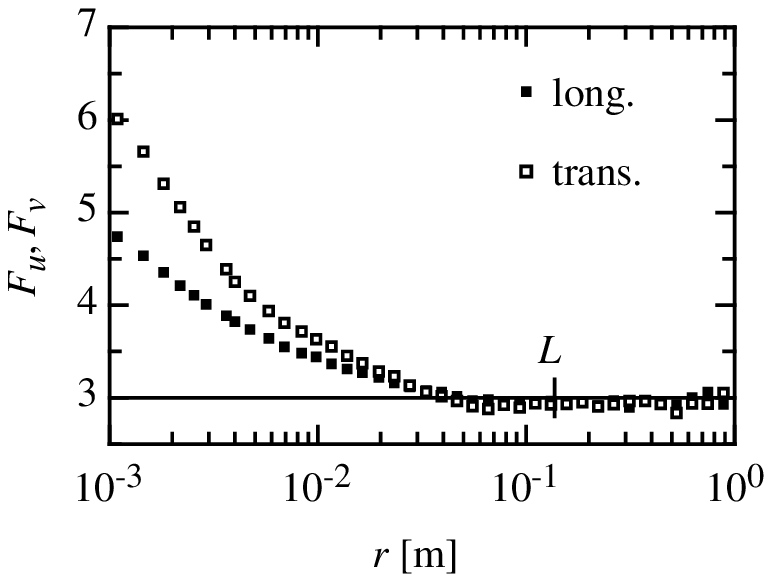}
\end{center}
\caption{\label{fig:flatness} Flatness of longitudinal and transverse increments. For scales $r<L$ the flatness lies for both components over the Gaussian value of 3. Furthermore the flatness is larger for the transverse component, i.e. it is more intermittent.}
\end{figure}

An other quantity which enables to quantify intermittency is the flatness $F_\alpha\equiv\left<\alpha^4\right>/\left<\alpha^2\right>^2$, where $\alpha$ is $u_r$ or $v_r$, respectively. For a Gaussian distribution it is $F_\alpha\equiv 3$ and for an intermittent distribution it is $F_\alpha> 3$. As shown in Fig. \ref{fig:flatness}, for $r<L$ both components are intermittent. For small $r$ the transverse increments are considerably more intermittent than the longitudinal one.

\begin{figure}[tbp]
\begin{center}
\includegraphics[width=5in]{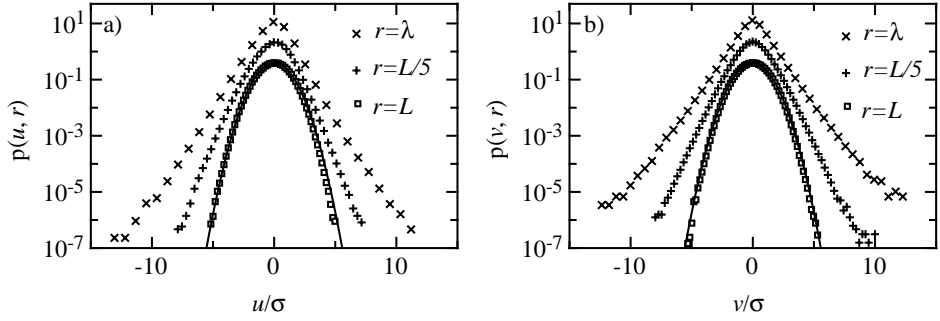}
\end{center}
\caption{Probability density functions (pdf) for a) longitudinal and b) transverse increments on three different length scales $r=L,\;L/5,\;\lambda\approx \lambda$. The pdfs' width are normalized with respect to their standard deviations and are shifted along the ordinate for a better representation. On the largest scale $L$ a Gaussian distribution is fitted for comparison. Towards smaller scales the deviations from the Gaussian form become obvious. \label{fig:pdf} }
\end{figure}

\begin{figure}[tbp]
\begin{center}
\includegraphics[width=5in]{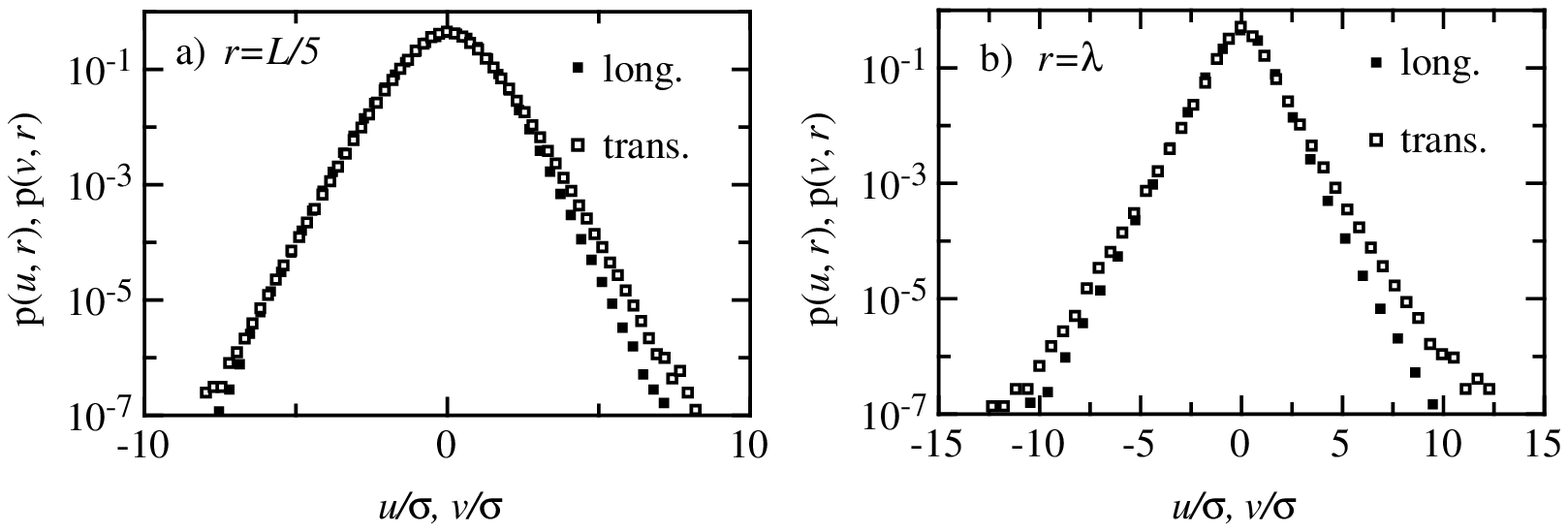}
\end{center}
\caption{A direct comparison between the longitudinal and transverse pdfs, a) for $r=L/5$ and b) for $r=\lambda$. Longitudinal pdfs are represented by black squares, transverse pdfs by white squares. The width of the pdfs are normalized with respect to their standard deviation. For $r=L/5$ the deviations are mainly given by the skewness, for $r=\lambda$ also differences in intermittency, i.e. exponential tails, occurs. \label{fig:pdf_uv} }
\end{figure}

In Fig. \ref{fig:pdf} the probability density of longitudinal and transversal velocity increments are plotted. The velocity distribution on each scale is normalized with the respective standard deviation, i.e. $u_{\sigma}:= u_r/\sigma_{u,r}$ and $v_{\sigma}:= v_r/\sigma_{v,r}$, to compare only the form of the curves. In Fig. \ref{fig:pdf_uv} the longitudinal and transversal probability densities for two different length scales ($r=L/5$ and $r=\lambda$) are shown. In both figures, the intermittent character of the statistics can be seen. Whereas for $L=L/5$ the main difference is seen only for positive increments, i.e. the main difference is the skewness, we find for smaller scales that the distributions for the transverse increments is for positive and negative increments more intermittent, see Fig. \ref{fig:pdf_uv}b).

\begin{figure}[tbp]
\begin{center}
\includegraphics[width=3.5in]{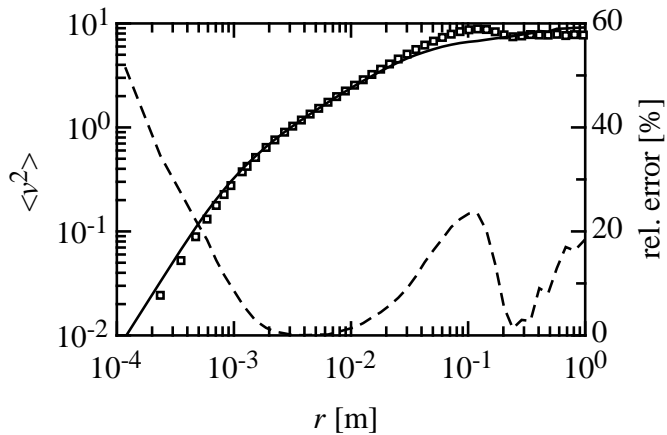}
\end{center}
\caption{The second order transverse structure function calculated from data (squares) and calculated from the longitudinal structure function via the K\'arm\'an equation (line). Because the differences in the scaling range can not be seen, also the relative error between them are plotted (dashed line). (R$_\lambda$=550)\label{fig:KarmanHRe}}
\end{figure}

\begin{figure}[tbp]
\begin{center}
\includegraphics[width=5in]{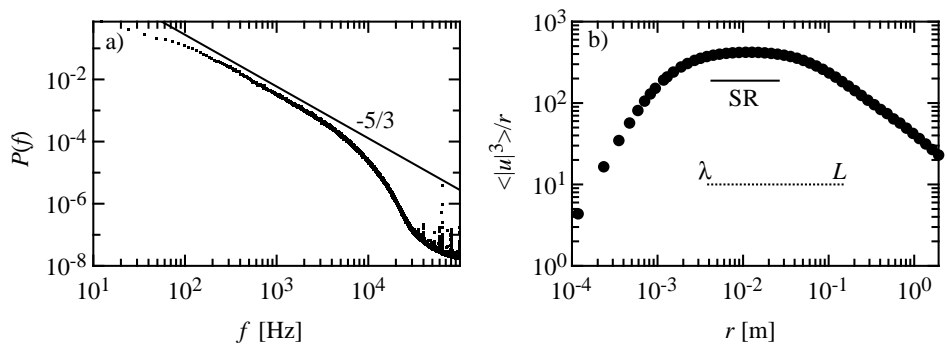}
\end{center}
\caption{a) Energy spectrum (dots) with a -5/3-power law and b) the third order structure function for R$_\lambda$=550. The plateau of $\left<|u|^3\right>/r$ defines the scaling range SR.\label{fig:scalingHRe}}
\end{figure}

Next we present briefly results from the high Reynolds number data set. The K\'arm\'an equation is well fulfilled, i.e. the data are isotropic in good approximation, see Fig. \ref{fig:KarmanHRe}. Fig. \ref{fig:scalingHRe}a) presents the energy spectrum of the longitudinal increments which shows a distinct scaling range with the exponent -5/3 in accordance to Kolmogorov's theory. The scaling range is more pronounced for this data set as it can be seen also from the third order structure function, compare Figs. \ref{fig:scalingHRe}b) and \ref{fig:scaling}b).  

In this section, we have presented our data with respect to two-point statistics, i.e. regarding only the statistics of increments for one fixed scale separately. The results are comparable to them given in the literature for moderate Reynolds numbers. Differences between longitudinal and transverse increments are clearly visible in pdfs and structure functions indicating that the transverse increments are more intermittent as the longitudinal. In the following we apply a new analysis to study the dependence between different length scale and the interaction between both increments which was not studied so far.

\subsection{Multi-Point Statistics: FP-Analysis}

In this section we present the analysis of the multi-point statistics separately for the longitudinal and transverse increments as it was described in section \ref{theory}. There are in essential three steps: First we show the validity of the Markov properties. Secondly, we calculate the Kramers-Moyal coefficients and show that the data obey a diffusion process. At last, as a verification, we integrate the resulting Fokker-Planck equation for the simple and for the conditioned probability distribution and show that the estimated Fokker-Planck equation describes correctly the data.

Inserting a comment on the increment definition: different from the usual increment definition, we define increments for the multi-point examinations according to $u_r:={\bf e}\cdot\left[ {\bf U}({\bf x}+{\bf r}/2)-{\bf U}({\bf x}-{\bf r}/2)\right]$. In appendix \ref{appendix}, we compare both definitions.

\subsubsection{Markov Properties}

\begin{figure}[tbp]
 \begin{center}
\includegraphics[width=5in]{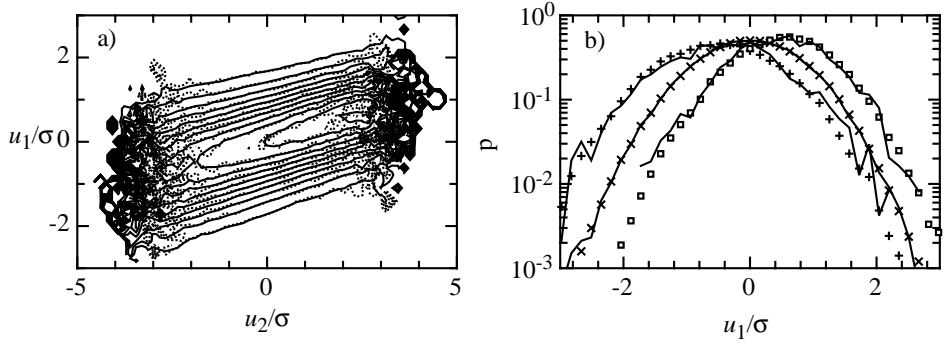}
\end{center}
\caption{a) Contour plot of the single (straight line) and double conditioned probability distribution (dashed line) $p(u_1,r_{1}|u_2,r_{2})$ and $p(u_1,r_{1}|u_2,r_{2};u_3\equiv 0,r_{3})$ of the longitudinal increments for the length scales $r_1=\Delta r$, $r_2=2\Delta r$ and $r_3=3\Delta r$ with $\Delta r=68.3\;$mm. The distance between the contour lines is $\Delta p = 0.05$. b) Three cuts through the contour plot are presented for $u_2=-2\sigma$, $u_2=0$ and $u_2=-2\sigma$. It can be seen that the Markov properties are fulfilled.\label{fig:markov1}}
\end{figure}

\begin{figure}[tbp]
\begin{center}
\includegraphics[width=5in]{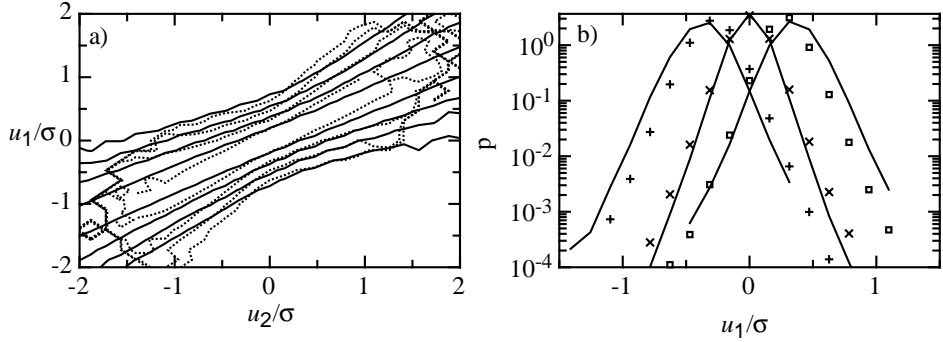}
\end{center}
\caption{The same plot as in Fig. \ref{fig:markov1} but for smaller values $\Delta r=2.54\;\textrm{mm}\approx \lambda/3$. a) Contour plot for the single conditioned (straight line) and double conditioned probability distribution (dashed line). Deviations between both are visible. The distance between the lines correspond to a factor 10. b) Three cuts through the contour plot are presented for $u_2=-2\sigma$, $u_2=0$ and $u_2=-2\sigma$. It can be seen that the Markov properties are not fulfilled.\label{fig:markov2}}
\end{figure}

The foundation of the following analysis is the validity of the Markov properties. They can be tested directly on data by their definition (\ref{MarkovCond}). Because of the finite numbers of measured data points we restrict ourself to the verification of $p(u_1,r_{1}|u_2,r_{2}) = p(u_1,r_{1}|u_2,r_{2};u_3,r_{3})$. Fig. \ref{fig:markov1} shows both side of this equation for the longitudinal increments for the three length scales $r_1=\Delta r$, $r_2=2\Delta r$ and $r_3=3\Delta r$ with $\Delta r=68.3\;$mm$\approx L/2$ . It can be seen that both distributions coincide. For a length scale below a certain threshold, the Markov properties are not fulfilled, see Fig. \ref{fig:markov2}. We call the associated length scale `Markov coherence length' or short `Markov length' $l_m$, i.e. above $l_m$ the Markovian properties are fulfilled, below $l_m$ not \cite{friedrich98}.
 
\begin{figure}[tbp]
\begin{center}
\includegraphics[width=5in]{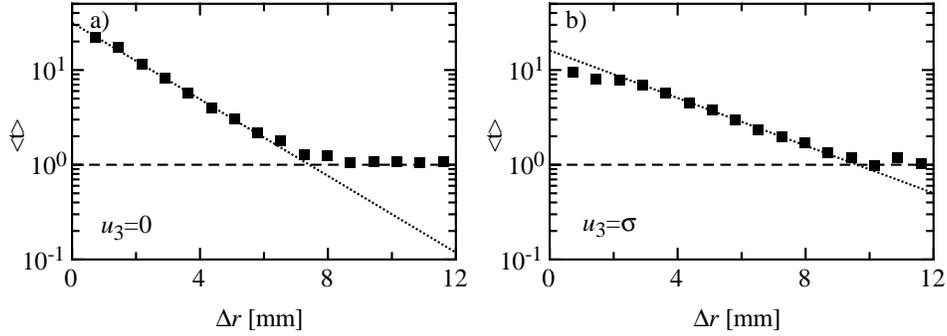}
\end{center}
\caption{The expectation value $\langle t\rangle$ of the Wilcoxon test in dependence of the length scale differences $\Delta r$ for the longitudinal increments with the reference scale $r_1=14.64\;$mm; a) for $u_3=0$ and b) for $u_3=\sigma$. The constant line marks the expectation value for fulfilled Markov properties, the dotted line is an exponential decay.\label{fig:markov3}}
\end{figure}

\begin{figure}[tbp]
\begin{center}
\includegraphics[width=5in]{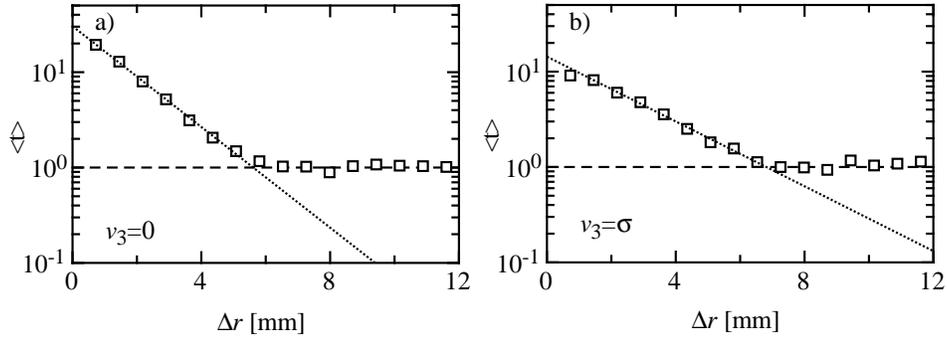}
\end{center}
\caption{ The same plot as in Fig. \ref{fig:markov3} but for the transverse increments. The only differences can be found in the smaller Markov-length. \label{fig:markov4}}
\end{figure}

To quantify the results and to get a more objective and systematic measure for the Markov properties and the Markov length, we perform a Wilcoxon test, which compares two random samples with size $m$ and $n$ (see \cite{wilcoxon45,renner01a}). For the Wilcoxon test, one has to count the number of inversions of two samples, here for the single and double conditioned variable $u_1|_{u_2}$ and $u_1|_{u_2,u_3}$. We calculate $\langle t\rangle:=|Q-\langle Q\rangle_{\tilde{p}=p}|/\sigma(m,n)$, where $Q$ is the number of inversions calculated from the experimental data for the variables $u_1|_{u_2}$ and $u_1|_{u_2,u_3}$; $\langle Q\rangle_{\tilde{p}=p}=mn/2$ and $\sigma(m,n)=\sqrt{mn(m+n+1)/12}$ are the number of inversions and the standard deviation, respectively, assuming that both variables have the same distribution. Thus, it is $\langle t\rangle=1$ if both samples come from the same universe, or have the same distribution. Fig. \ref{fig:markov3} shows this measurement in dependence of $\Delta r$. For small $\Delta r$ the Markov properties are not fulfilled, whereas for larger distances the deviations are not significant anymore. We identify the distance $\Delta r=l_{m}$, where $\langle t\rangle$ drops to 1 as the Markov length \footnote{We have also performed the Kolmogorov-Smirnov test with similar results, but for our purpose the Wilcoxon test seems to be more sensitive.}. This value can be estimated by fitting an exponential function to the values and by interpreting the passage through the value 1 as the Markov length, as it is presented in Fig. \ref{fig:markov3}. The Markov properties are also fulfilled for transverse increments, but with a smaller Markov length, see Fig. \ref{fig:markov4}. The Markov length varies within 20\% with respect to the condition $u_2$ but remains about constant with respect to $r$. For the longitudinal increments the Markov length lies in the range $7.4\,\textrm{mm}<l_{m,l}<9.6\,\textrm{mm}$,  for the transversal increments the Markov length lies in the range $5.5\,\textrm{mm}<l_{m,t}<6.8\,\textrm{mm}$.  The ratio is $l_{m,l}/l_{m,t}\approx 1.4$ as it is known for the Taylor length \cite{Pope00}. Note, that up to now wee see that $l_{m}\approx \lambda$ \cite{renner01a}. The analysis of the data R$\lambda$=550 give in principle analogous results.


We conclude from the results that the `cascade' of longitudinal and transverse increments can be described by a Markov-process for stepsizes larger than the Markov length $l_m$, which becomes important again for the estimation of the Kramers-Moyal coefficients, as will be seen below.

\subsubsection{Kramers-Moyal Coefficients}

\begin{figure}[tbp]
\begin{center}
\includegraphics[width=5in]{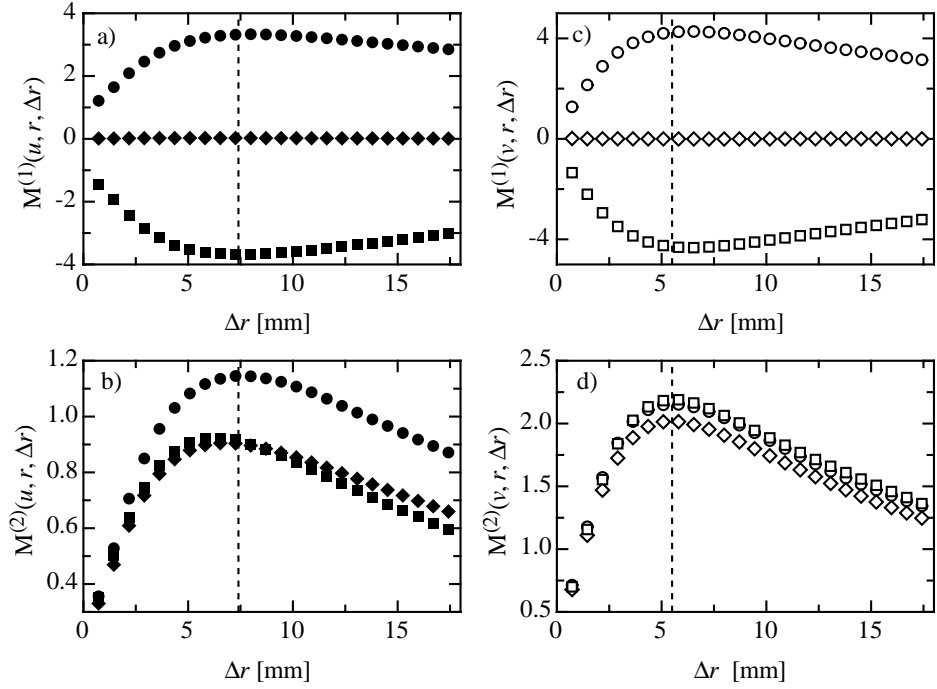}
\end{center}
\caption{The coefficients $M^{(1)}$ and $M^{(2)}$ in dependence of the step-size $\Delta r$ for $r=L/2$. The two upper figures show the limit of the drift coefficient for the longitudinal and transverse increments, respectively. The two lower figures show the limit of the diffusion coefficient for the longitudinal and transverse increments. The dotted lines represent the Markov length. Circles: $u,\,v=-2\sigma$, diamonts: $u,\,v=0$, squares: $u,\,v=2\sigma$.\label{fig:limes}}
\end{figure}

The drift coefficients $D^{(1)}$ and the diffusion coefficients $D^{(2)}$ are calculated according to Eq. (\ref{DkDef}) directly from the measured data following the procedure described in \cite{renner01a,friedrich02}. The crucial point is the estimation of the limit $\lim_{\Delta r\to 0}M^{(i)}$, see Fig. \ref{fig:limes}. Only the points with $\Delta r>l_\textrm{m}$ are used to estimate the limit. For $\Delta r<l_\textrm{m}$ the Markov-properties are violated, for $\Delta r>l_\textrm{m}$ the dependence of $M^{(i)}$ with respect to $\Delta r$ is linear, see Fig. \ref{fig:limes}. We thus use a first order polynomial to extrapolate the limit $\Delta r\to 0$. The linear dependence is the first order approximation of the limit, see \cite{friedrich02}.

\begin{figure}[tbp]
\begin{center}
\includegraphics[width=5in]{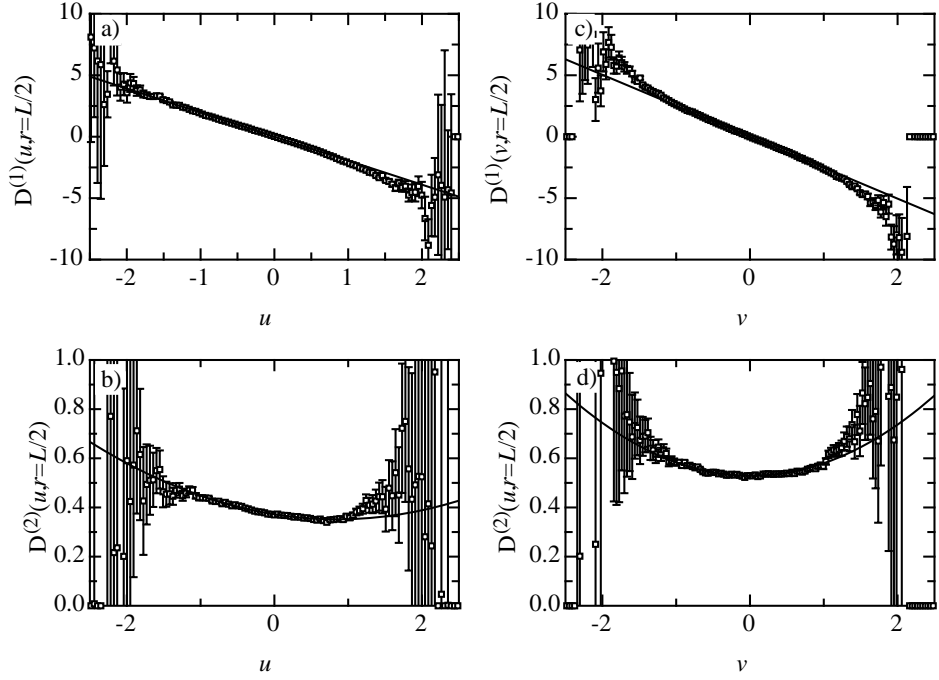}
\end{center}
\caption{The Kramers-Moyal coefficients $D^{(1)}$ and $D^{(2)}$ and their dependence of the increments $u$ and $v$, respectively for $r=L/2$. a) and b) longitudinal, c) and d) transverse increments. \label{fig:uDependence}}
\end{figure}

The resulting drift coefficients $D^{(1)}$ and diffusion coefficients $D^{(2)}$ are shown in Fig. \ref{fig:uDependence} for the length scale $r=L/2$. The drift coefficient can be approximated by a straight line with negative slope. Small deviations from this behavior are visible for the transverse component. For the diffusion coefficient qualitative differences between both increments are visible. In contrast to the transverse coefficient, the longitudinal coefficient is not symmetric under reflection $u\to -u$. This is compatible with Kolmogorov's 4/5-law, which states that the longitudinal distribution are skewed. The diffusion coefficient can be approximated by a second order polynomial, so we have
\begin{eqnarray}\label{ds1}
D^{(1)}(\alpha,r)&=&d^\alpha_1(r)\alpha\\\nonumber
D^{(2)}(\alpha,r)&=&d_{2}(r)+d^\alpha_{2}(r)\alpha+d^{\alpha\alpha}_{2}(r)\alpha^2,\\\nonumber
\end{eqnarray}
where $\alpha=u,v$. Due to the reflection symmetry $v\to -v$ of the transverse increments it is $d^v_{2}(r)\equiv 0$.

\begin{figure}[tbp]
\begin{center}
\includegraphics[width=5in]{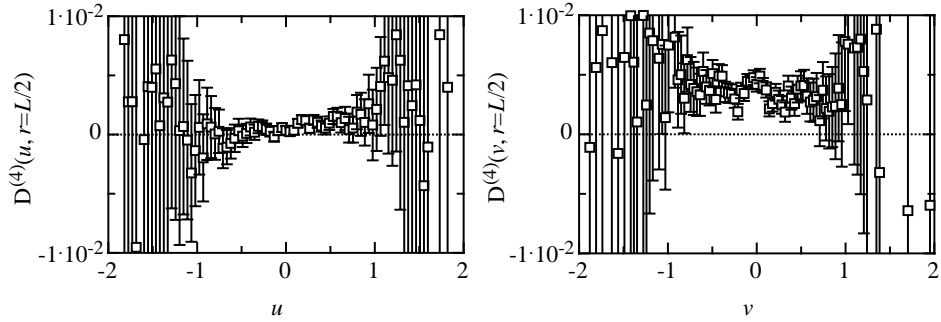}
\end{center}
\caption{The fourth order Kramers-Moyal coefficients $D^{(4)}(u,r)$ and $D^{(4)}(v,r)$ for $r=L/2$. It can be seen that within the error-bars the longitudinal coefficient is zero. The transversal coefficient is slightly above zero, but its contribution to the Kramers-Moyal expansion is small.\label{fig:M4}}
\end{figure}

The drift and diffusion coefficients are the first two coefficients of the Kramers-Moyal expansion. According to the Pawula-theorem all higher coefficients vanish, if the fourth-order coefficients are zero and thus the expansion simplifies to a Fokker-Planck equation. In Fig. \ref{fig:M4} this fourth order coefficients are plotted for $r=L/2$. The coefficient $D^{(4)}$ for the longitudinal coefficient vanish within the error bars. The corresponding transversal coefficient has a value slightly above zero. But one can estimate with the Kramers-Moyal expansion that the contribution of this coefficient is less than one-hundredth in comparison to the diffusion coefficient. Therefore, we want to assume in the following a vanishing fourth order coefficient. This assumption is additionally justified below by showing that the Fokker-Planck equation describes the increment statistics well.

\begin{figure}[tbp]
\begin{center}
\includegraphics[width=5in]{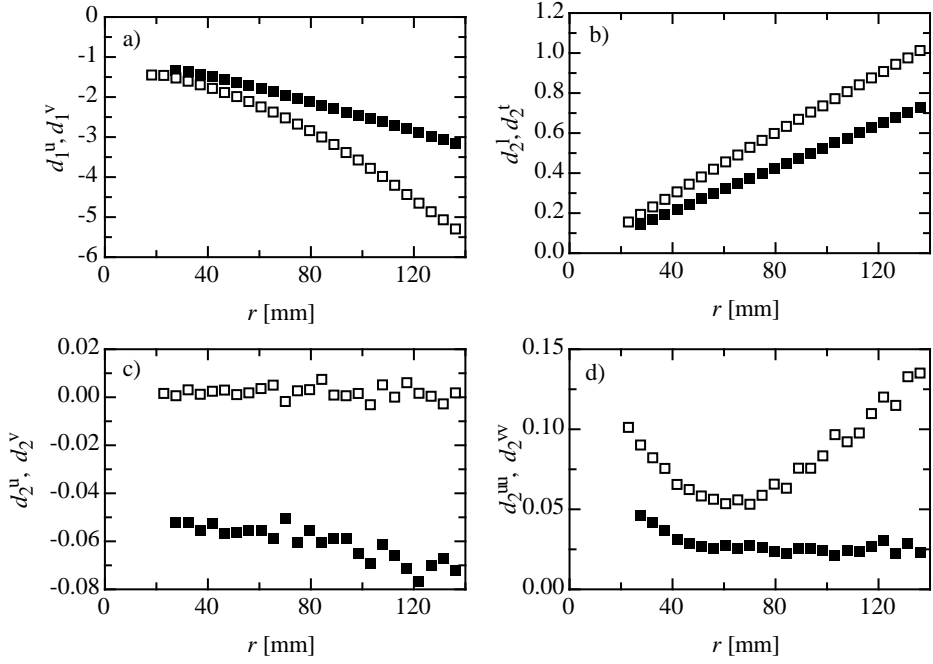}
\end{center}
\caption{The $r$-dependence of the expansion coefficients of the Kramers-Moyal coefficients according to Eq. (\ref{ds1}) for the longitudinal (black squares; $\alpha=u$) and transverse increments (white squares; $\alpha=v$) If the notation is unique, we omit the index $l$ or $t$. \label{fig:CoeffCenterNorm}}
\end{figure}

Next, the $r$-dependence in Eq. (\ref{ds1}) is investigated. It can be estimated by fitting the approximation (\ref{ds1}) to the numerical Kramers-Moyal coefficients. The result is depicted in Fig. \ref{fig:CoeffCenterNorm} for the longitudinal (black squares) and transverse increments (white squares). The form is remarkably simple, it can be approximated by
\begin{eqnarray}\label{ds1d}
&&D^{(1)}_\textrm{l}:\;\;\;\;d^u_{1,\textrm{l}}(r)=\alpha^u_{1,\textrm{l}}+\beta^u_{1,\textrm{l}}r\;\;\;\\\nonumber
&&D^{(1)}_\textrm{t}:\;\;\;\;d^v_{1,\textrm{t}}(r)=\alpha^v_{1,\textrm{t}}+\beta^v_{1,\textrm{t}}r+\gamma^v_{1,\textrm{t}}r^2\\\nonumber
&&D^{(2)}_\textrm{l}:\;\;\;\;d_{2,\textrm{l}}(r)=\beta_{2,\textrm{l}}r,\;\;\;d^u_{2,\textrm{l}}(r)=\beta^u_{2,\textrm{l}}r,\;\;\;d^{uu}_{2,\textrm{l}}(r)=\alpha^{uu}_{2,\textrm{l}}\\\nonumber
&&D^{(2)}_\textrm{t}:\;\;\;\;d_{2,\textrm{t}}(r)=\beta_{2,\textrm{t}}r,\;\;\;
d^{vv}_{2,\textrm{t}}(r)=\alpha^{vv}_{2,\textrm{t}}+\beta^{vv}_{2,\textrm{t}}r+\gamma^{vv}_{2,\textrm{t}}r^2.\\\nonumber
\end{eqnarray}
Here, we denote by $X_l$ and $X_t$ the longitudinal and transverse quantities, respectively.

\subsubsection{Integration of the Fokker-Planck Equation}

\begin{figure}[tbp]
\begin{center}
\includegraphics[width=2.5in]{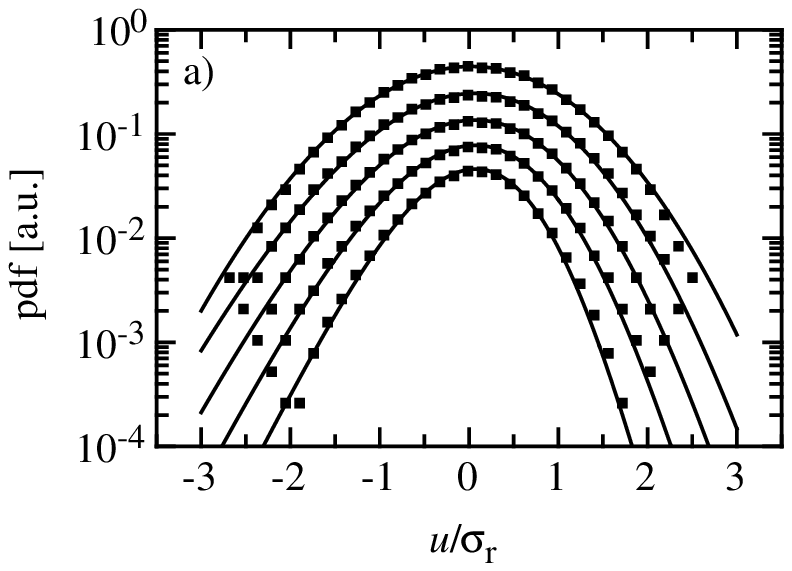}
\includegraphics[width=2.5in]{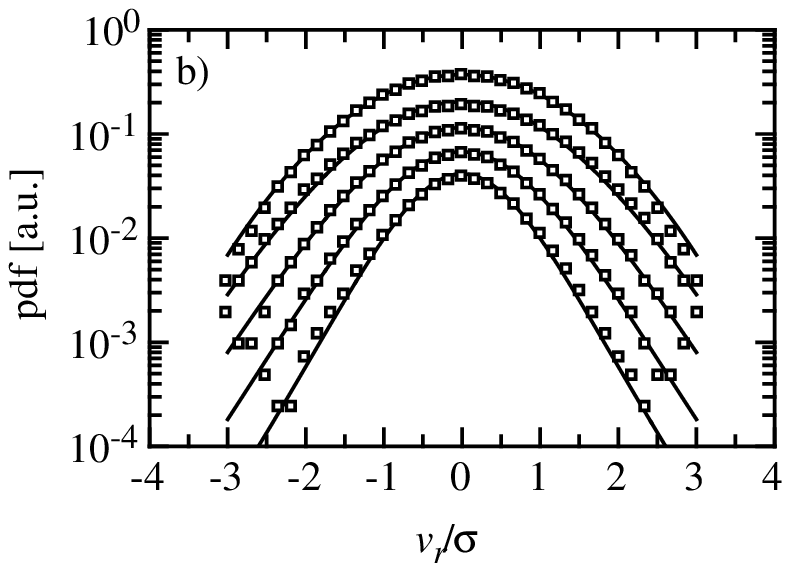}
\end{center}
\caption{The solution of the Fokker-Planck equation in comparison to the data. With the initial distribution on the scale $r=L$ the Fokker-Planck equation is solved numericaly. The curves belongs to the scales  $r=$131, 74, 41, 23, 12$\;$mm from top to bottom. For a better visibility the curves are shifted by a constant factor. a) The solution for longitudinal and b) for transverse increments. In both pictures the intermittency is visible, for the longitudinal increments also the skewness can be seen.\label{fig:fopla}}
\end{figure}

\begin{figure}[tbp]
\begin{center}
\includegraphics[width=2.5in]{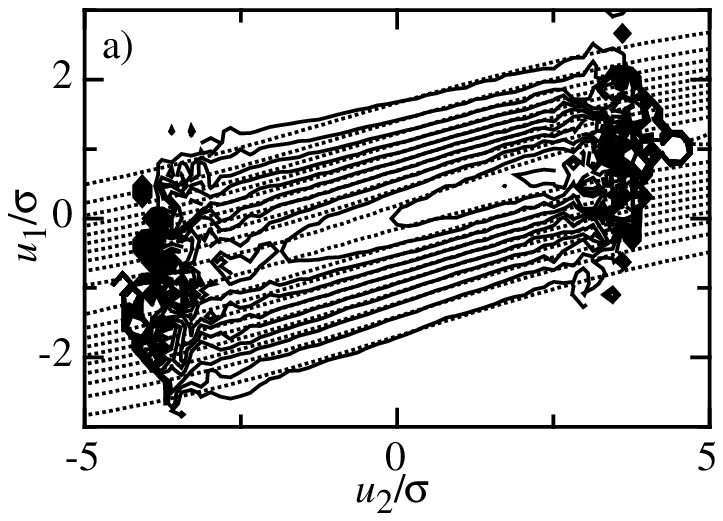}
\includegraphics[width=2.5in]{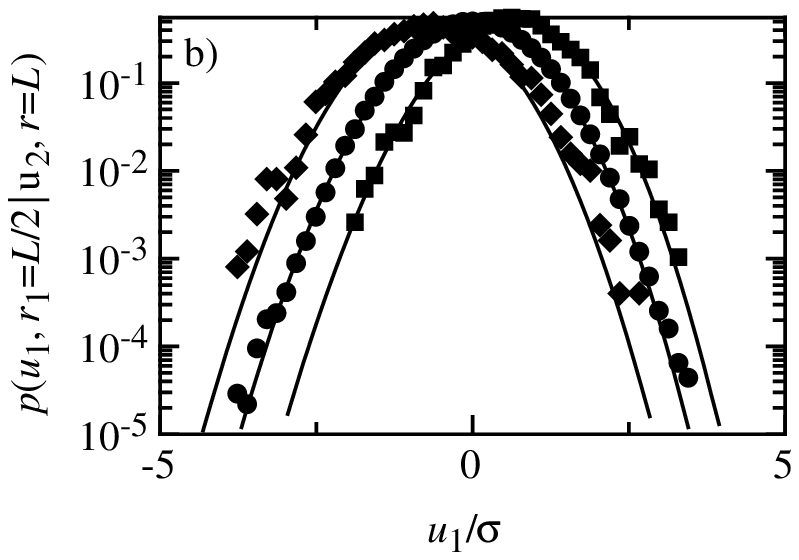}
\end{center}
\caption{The longitudinal conditional pdfs $p(u_2,r=L/2|u_1,r=L)$ calculated from $r=L$ down to $r=L/2$. a) The contour plot for the numerical solution (dashed line) and pdfs calculated from the data (straight line). The distance between the contour lines is $\Delta p = 0.05$. b) Cuts through the contour plot for  $u_2=-2.5\sigma,\;0,\;2.5\sigma$. The lines are the solution of the Fokker-Planck equation, the symbols represent the pdfs from the data.\label{fig:trans1}}
\end{figure}

\begin{figure}[tbp]
\begin{center}
\includegraphics[width=2.5in]{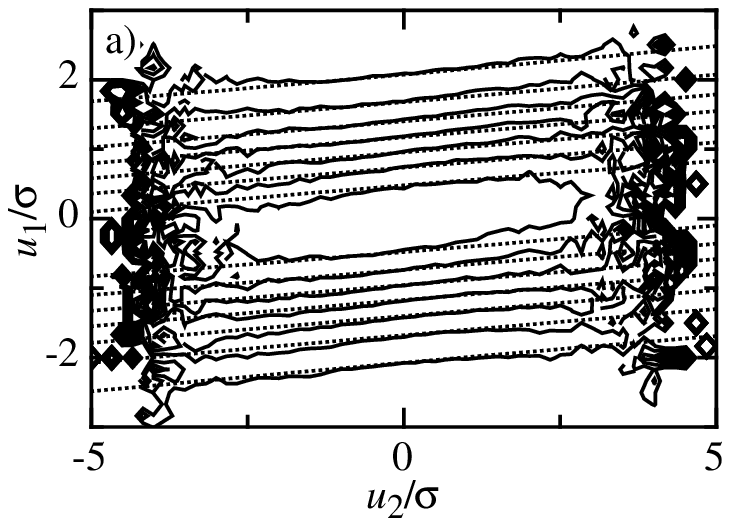}
\includegraphics[width=2.5in]{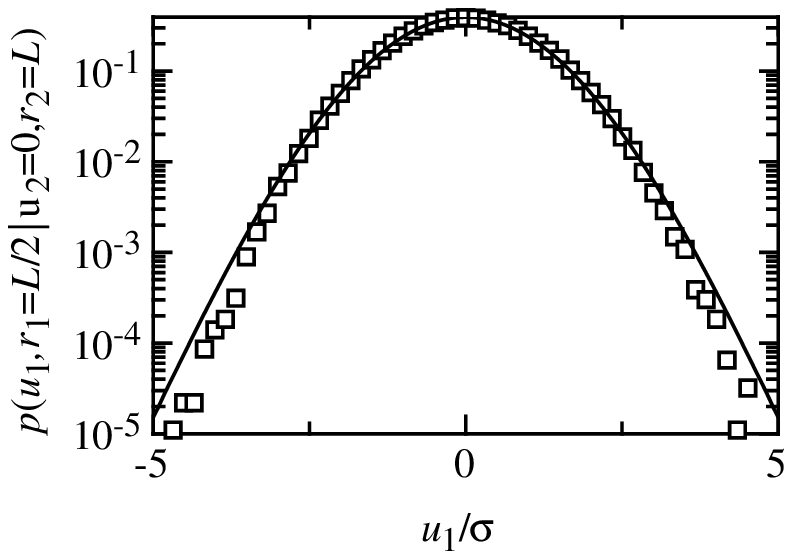}
\end{center}
\caption{The same as Fig. \ref{fig:trans1} but for the transverse increments. a) The contour plot for the numerical solution (dashed line) and pdfs calculated from the data (straight line). The distance between the contour lines is $\Delta p = 0.05$. b) Cuts through the contour plot for  $u_2=0$. The line is the solution of the Fokker-Planck equation, the symbols represent the pdf from the data.\label{fig:trans2}}
\end{figure}

Next we want to demonstrate that the Fokker-Planck equation can describe correctly the statistics of the turbulent field. If the estimated drift and diffusion coefficients (\ref{ds1d}) are used to solve the Fokker-Planck equation numerically, the resulting distributions can be compared with the distributions estimated directly from the data. First, the numerical estimation of the single distributions $p(u,r)$ and $p(v,r)$ in dependence of $r$ are discussed, for which the distribution on the integral scale $p(u,r=L)$ and $p(v,r=L)$ are used as initial conditions. In Fig. \ref{fig:fopla} $p(u,r)$ and $p(v,r)$ are shown for several length scales. The curves are in good agreement with the data. It is important to stress that also the intermittency effects and the skewness can be described well. A similar calculation has been done for the conditional distributions $p(u,r|u_0,r_0)$ and $p(v,r|v_0,r_0)$ starting at the integral scale with Dirac's delta function $p(u,r=L|u_0,r_0=L)=\delta(u)$ and $p(v,r=L|v_0,r_0=L)=\delta(v)$ as initial condition. The solutions down to $r=L/2$ are shown in Figs. \ref{fig:trans1} and \ref{fig:trans2} in dependence of the initial value $u_0$ and $v_0$.
Comparing the conditional probability distributions in Fig. \ref{fig:trans1} with those in Fig. \ref{fig:trans2} it is evident that the transverse statistics relaxes faster (the contour lines are more horizontal).

On the basis of these results we conclude that the increment statistics can be well described by a Fokker-Planck equation, whose drift and diffusion coefficients are given by (\ref{ds1d}). Thus, we can examine the increment statistics by means of the drift coefficient $D^{(1)}$ and diffusion coefficient  $D^{(2)}$.

\begin{figure}[tbp]
\begin{center}
\includegraphics[width=5in]{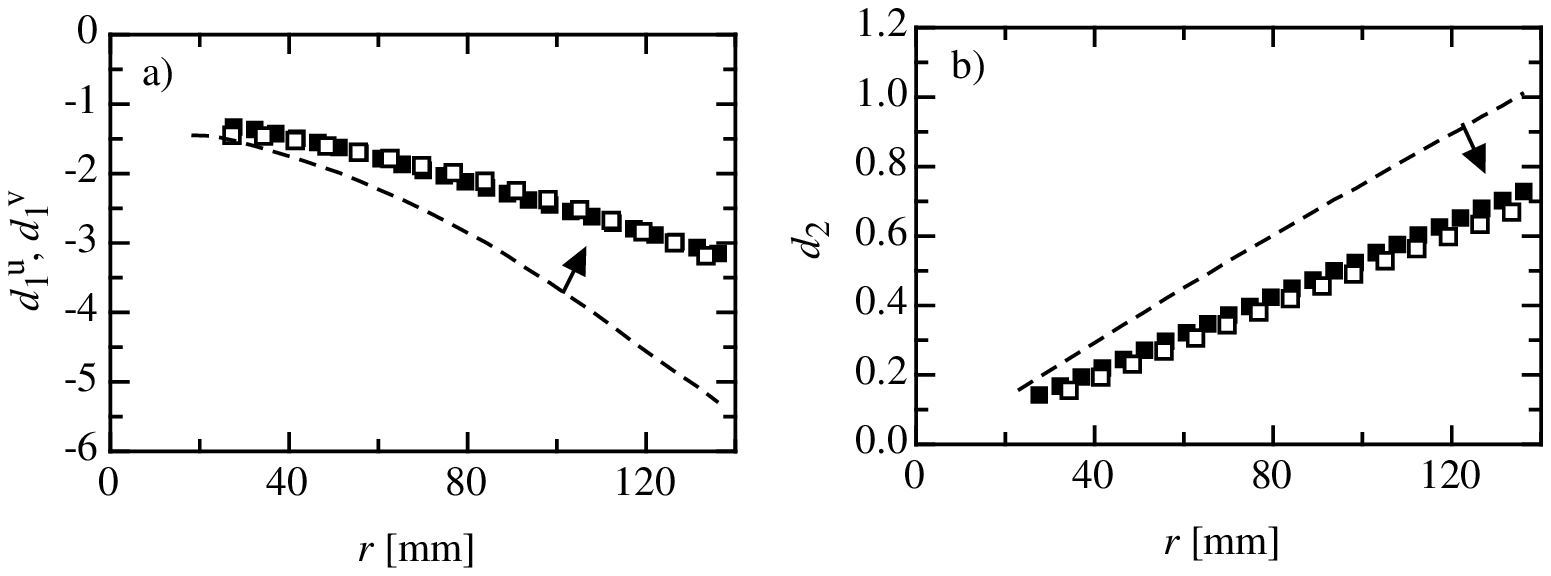}
\end{center}
\caption{The drift coefficient a) and the constant term of diffusion coefficient b) of the longitudinal (black squares) and transverse component (interrupted line) can be shifted on top of the other by stretching the $r$-dependence of the transverse coefficients by a factor 2/3 (white squares).\label{fig:2_3_1D}}
\end{figure}

\begin{figure}[tbp]
\begin{center}
\includegraphics[width=3.5in]{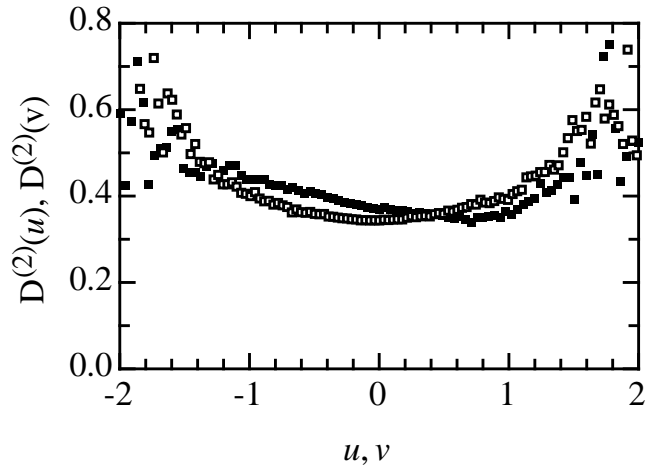}
\end{center}
\caption{The diffusion coefficient of the longitudinal increments on $r=L/2$ (black squares) and the transverse increments on $r=2/3\cdot L/2$ (white squares). It can be seen the minima for both are the same.\label{fig:2_3_D2}}
\end{figure}

A closer look at the $d$-coefficients of Eq. (\ref{ds1d}) shown in Fig. \ref{fig:CoeffCenterNorm} gives insight into the differences of the longitudinal and transverse increment statistic. It can be seen in Fig. \ref{fig:CoeffCenterNorm}a) b) and d) that the absolute value of the transverse coefficients are larger than the longitudinal one. This means that the transverse cascade are in a way `faster' and more noisy. This behavior can be state more precisely by taking into account a remarkably symmetry, namely, if the $r$-dependence of the transverse increments is rescaled by the factor 2/3, i.e. $r\to 2r/3$, the dominating terms fall on top of the others, see Fig. \ref{fig:2_3_1D}. Thus the main statistical differences between longitudinal and transverse increments vanish by this rescaling. Differences only remains in the diffusion term, see Fig. \ref{fig:2_3_D2}. In section \ref{sec:discussion} we discuss this result in further detail.


\subsection[Joint statistics of longitudinal and transverse increments]{Joint multipoint-statistics of longitudinal and transverse increments}
\label{sec:2d}

In the preceding section we have analyzed the statistics of longitudinal and transverse increments separately. But this separation is restrictive because the dynamics of both components come from one velocity field and are therefore connected. They can not be separated due to the nonlinear advection term in the Navier-Stokes equation. Therefore we extend the above analysis and examine the joint stochastic properties.  We chose the both increments as one state with the scale parameter $r$, see Eq. (\ref{ustate}). The aim is to estimate a Fokker-Planck equation (\ref{FoplaCond}) in these two variables. We proceed similar to the one dimension analysis. The main difference is that one need much more data and one has to estimate much more coefficients, because the drift-coefficient is now a vector and the diffusion coefficient a matrix.

The verification of the Markov properties for two variables is doubtful because one has to estimate the double conditioned probability function for a two dimensional process, i.e. a six-dimensional function with a finite number of data points. One needs approximately $10^4$ times more data points in comparison to the one dimensional case if the results should be similar significant. The limiting factor is the duration of the measurement and the amount of data. But we know separately for both components that the the Markov properties are valid. In general, if two variables have Markov properties, also the joint statistics have Markov properties (the reverse is not true in general \cite{Risken}). Therefore we assume that the combined process is Markovian as well.

The next step is to estimate the Kramers-Moyal coefficients. First, one has to calculate the approximation of the drift vector $M^{(1)}_{i}(u,v,r,\Delta r)$ and of the diffusion matrix $M^{(2)}_{ij}(u,v,r,\Delta r)$ in dependence of $\Delta r$, see Eq. (\ref{MkDef}).
 As in the one dimensional case we calculate the approximation of the drift and diffusion coefficients by fitting a linear polynomial to these functions in dependence of $\Delta r$ above the Markov length. In doing so, we have to consider two different Markov length for both components, one for each component.

\begin{figure}[tbp]
\begin{center}
\includegraphics[width=3.5in]{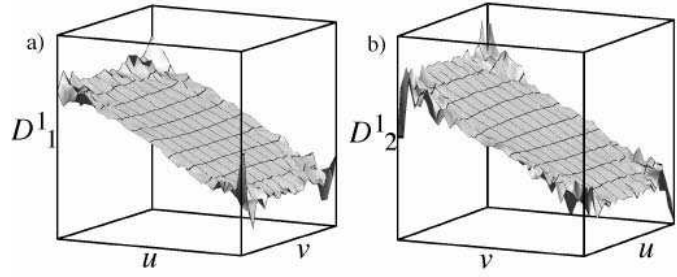}
\end{center}
\caption{The $u$ and $v$ dependence of the drift vector for the scale $r=L/4$. a) The drift coefficient $D^{(1)}_1$, b) the drift coefficient $D^{(1)}_2$. Note that the vertical axis is rotated for an better comparison between a) and b). Both coefficients are pure linear and depend only on one variable.\label{fig:D1_3D} }
\end{figure}

\begin{figure}[tbp]
\begin{center}
\includegraphics[width=5in]{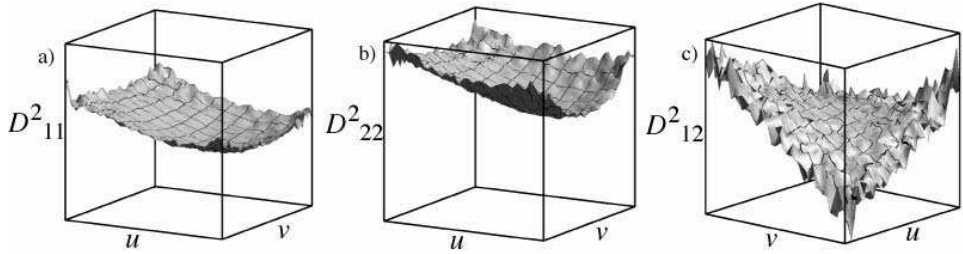}
\end{center}
\caption{The $u$ and $v$-dependence of the diffusion matrix for the scale $r=L/4$. a) The coefficient $D^{(2)}_{11}$, b) the coefficient $D^{(2)}_{22}$. It can be seen that the diagonal coefficient are not constant but have a parabolic form, which is more pronounced for the $D^{(2)}_{22}$ coefficient (multiplicative noise). Both coefficients are symmetric under reflection with respect to $v\to -v$, but not for $u\to -u$. c) The saddle-formed off-diagonal coefficient. \label{fig:D2_3D} }
\end{figure}

The shape of the drift  and diffusion coeffients are shown in Figs. \ref{fig:D1_3D} and \ref{fig:D2_3D} for the length scale $L/4$. The drift coefficients have a simple form: they depend only linearly on one variable, see Fig. \ref{fig:D1_3D}. That means the process decouples in the deterministic part. Remark if this was not the case the increments could not be Markovian seperately in contradiction to our findings. The diagonal coefficients of the diffusion matrix are shown in Figs. \ref{fig:D2_3D}a) and \ref{fig:D2_3D}b). Roughly spoken they consist of a curved surface shifted upwards with a minimum at $u>0$ and $v=0$. The off-diagonal coefficient $D^{(2)}_{12}$ (it is $D^{(2)}_{12}\equiv D^{(2)}_{21}$) has a saddle shape, see Fig. \ref{fig:D2_3D}c).

To summarize, we can approximate the form of the drift and diffusion coefficients by a low-order polynomial in $u$ and $v$:
\begin{eqnarray}\label{ds}
D^{(1)}_1(u,v,r)&=&d^u_1(r)u\\\nonumber
D^{(1)}_2(u,v,r)&=&d^v_2(r)v\\\nonumber
D^{(2)}_{11}(u,v,r)&=&d_{11}(r)+d^u_{11}(r)u+d^{uu}_{11}(r)u^2+d^{vv}_{11}(r)v^2\\\nonumber
D^{(2)}_{22}(u,v,r)&=&d_{22}(r)+d^u_{22}(r)u+d^{uu}_{22}(r)u^2+d^{vv}_{22}(r)v^2\\\nonumber
D^{(2)}_{12}(u,v,r)&=&d_{12}(r)+d^v_{12}(r)v+d^{uv}_{12}(r)uv\nonumber.
\end{eqnarray}
The $d$-coefficients contain the $r$-dependence. Their lower index labels the associated Kramers-Moyal coefficient, the upper index the order of the coefficient with respect to $u$ and $v$. In the Fokker-Planck equation, the coefficients occur symmetric with respect to reflection $v\to -v$. Thus a reflection $v\to -v$ does not change anything, whereas this symmetry is violated for the longitudinal increment. Of course, the Kramers-Moyal coefficients can be better approximated using higher order in $u$ and $v$, but their contributions are small and their value are not well defined at the edges of the available data range, because high velocities are too rare to ensure a good statistics. Note, the significance of higher order terms in $D^{(1)}$ and $D^{(2)}$ is important for the closure of Eqn. (\ref{StrukturGl}) but their investigation should not be addressed in this article.

\begin{figure}[tbp]
\begin{center}
\includegraphics[width=3.5in]{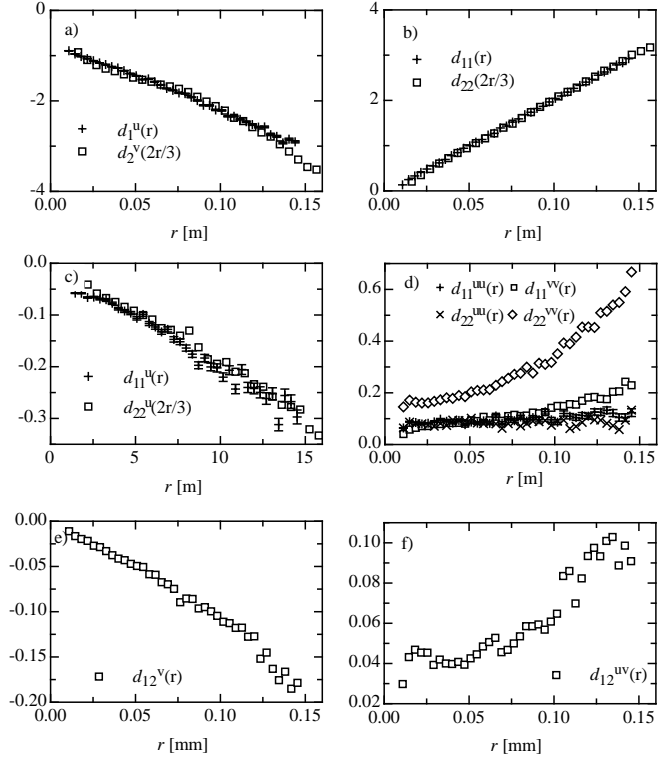}
\end{center}
\caption{The coefficients $d$ of Eq. (\ref{ds}), which reflect the dependence of the Kramers-Moyal equation on $u$ and $v$ in dependence of the linear scale variable  $r$. In Figs. a), b) and c), the argument of the transversal coefficient is rescaled by the factor 2/3. The corresponding longitudinal and transversal coefficients coincides which each other apart from the intermittency term $d_{22}^{vv}$.  e) and f): the coefficients for the off-diagonal diffusion coefficients.
\label{fig:d_r} }
\end{figure}

In order to describe the statistics with the Fokker-Planck equation completely, the $r$-dependence of the $d$-coefficients has to be estimated, see Fig. \ref{fig:d_r}. It can be approximated by
\begin{eqnarray}\label{d_r}
&&D^{(1)}_1:\;\;\;\;d^u_1(r)=\alpha^u_1+\beta^u_1r\;\;\;\\\nonumber
&&D^{(1)}_2:\;\;\;\;d^v_1(r)=\alpha^v_1+\beta^v_1r+\gamma^v_1r^2\\\nonumber
&&D^{(2)}_{11}:\;\;\;\;d_{11}(r)=\beta_{11}r,\;\;\;d^u_{11}(r)=\beta^u_{11}r,\;\;\;d^{uu}_{11}(r)=\alpha^{uu}_{11},\;\;\;d^{vv}_{11}(r)=\alpha^{vv}_{11}\\\nonumber
&&D^{(2)}_{22}:\;\;\;\;d_{22}(r)=\beta_{22}r,\;\;\;
d^u_{22}(r)=\beta^u_{22}r,\;\;\;d^{uu}_{22}(r)=\alpha^{uu}_{22},\;\;\;\\&&\;\;\;\;\;\;\;\;\;\;\;\;\;\,
d^{vv}_{22}(r)=\alpha^{vv}_{22}+\beta^{vv}_{22}r+\gamma^{vv}_{22}r^2\\\nonumber
&&D^{(2)}_{12}:\;\;\;\;d_{12}^v(r)=\beta_{12}^vr,\;\;\;\alpha^{uv}_{12}+\beta^{uv}_{12}r+\gamma^{uv}_{12}r^2\nonumber.
\end{eqnarray}

\begin{figure}[tbp]
\begin{center}
\includegraphics[width=3.5in]{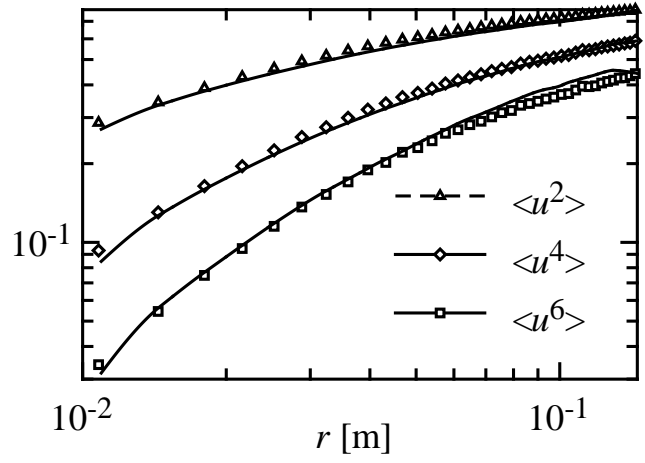}
\end{center}
\caption{\label{fig:verific}The even longitudinal structure functions up to order 6 calculated directly from data (symbols) compared to that of the Fokker-Planck equation using Eq. (\ref{StrukturGl}) (solid line).}
\end{figure}

\begin{figure}[tbp]
\begin{center}
\includegraphics[width=5in]{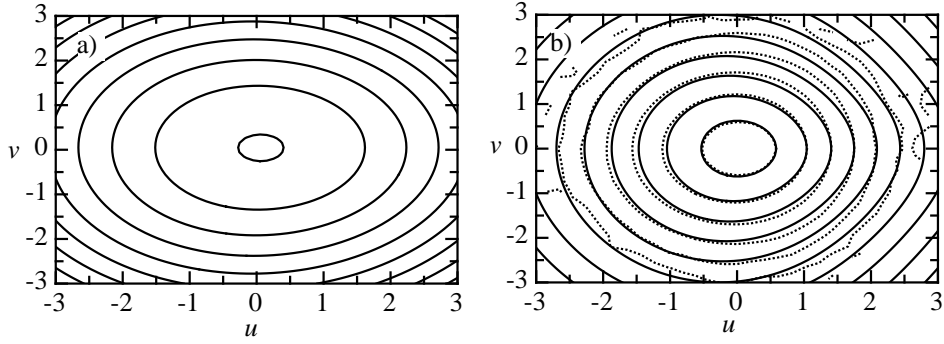}
\end{center}
\caption{Solution of the Fokker-Planck equation. a) Contour plot of the initial condition in logarithmic scale. The simulation starts at the integral length $r=L$ with a Gaussian distribution fitted to the data. b) The contour plots in logarithm scale of the simulated probability distribution on the scale $r=2\lambda$. The distance between the contour lines is chosen in logarithmic scale and correspond to a factor 10. Dashed lines are the probability calculated directly from data, the full lines are the simulation ones. The simulation reproduces well the properties of the data.\label{fig:verific2}}
\end{figure}

\begin{figure}[tbp]
\begin{center}
\includegraphics[width=5in]{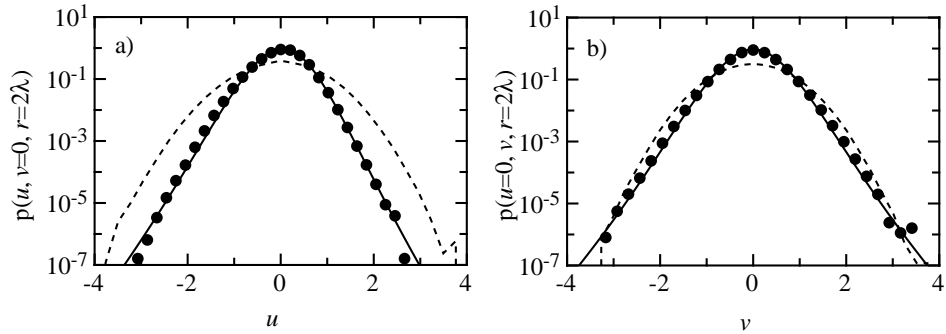}
\end{center}
\caption{Cuts through the probability distributions shown in Fig. \ref{fig:verific2}. Black circles are the pdfs of the data, the straight line is the simulation and the dashed line is the initial condition at $r=L$. a) The distribution in dependence of $u$ for $v=0$. b) The distribution in dependence of $v$ for $u=0$. Notice that the skewness in Fig. a) and the intermittency for both components are well reproduced.\label{fig:verific3}}
\end{figure}

Before we interpret the results, it has to be shown that the Fokker-Planck equation together with the coefficients (\ref{ds}) and (\ref{d_r}) can reproduce the statistics of the measured data. There are in principal two ways to verify the estimated drift and diffusion coefficients. One way is to solve the Fokker-Planck equation, the other way is to calculate the structure functions, which can be done for example with the hierarchical equation for the structure functions (\ref{StrukturGl}). In both cases the results can be compared with the corresponding quantities estimated directly from the data. In Fig. \ref{fig:verific} the solution of the hierarchical structure function equation (\ref{StrukturGl}) is shown for $n=0$ and $m=1,\,\dots,6$. It is in good agreement with the structure functions. To show that also the joint probability distributions can be reproduced, we solve the Fokker-Planck equation by calculating the path-integral (\ref{pathint}) numerically. We use the estimated Kramers-Moyal coefficients and start the simulation on the integral scale $r=L$ with a Gaussian distribution for $p(\mathbf{u}_0,r=L)$ and integrate down to $r=2\lambda$. Figs. \ref{fig:verific2} and \ref{fig:verific3} show the results. Because of the good accordance we assume that the $d$-coefficients can be used to characterize the statistics of longitudinal and transverse increments.


The number of coefficients in (\ref{d_r}) can be reduced by a remarkably symmetry. If one multiplies the $r$-dependence for the transverse increments with a factor $2/3$, the related coefficients of longitudinal and transverse increments coincides, i.e. $d_\textrm{long}(r)\approx d_\textrm{trans}(2r/3)$. The only exception is the diffusion term $d^{vv}_{22}(r)$. Thus the number of independent coefficients is halved.
This symmetry will be examined in more detail below.

\section{Interpretation of the Results}
\label{sec:discussion}

From the above analysis we know that the drift and diffusion coefficients contain the information of the small-scale statistics of the turbulence. More specifically, they contain the joint statistics of longitudinal and transverse increments as well as the statistics on multi scales. Thus we can study the structure of small-scale turbulence with aid of the drift and diffusion coefficients.

First we will argue that the hierarchical equation (\ref{StrukturGl}) is a kind of generalized K\'arm\'an equation and we discuss the similarities and differences between these two equations. Then we consider the 2/3-rescaling symmetry in more detail, show that it is consistent with known results and find that the coupling of longitudinal and transversal increments is important for intermittency. 
 At least we show that the scaling exponents of the transverse structure function can fake a too small value if the frequently used extended self similarity (ESS) is used to estimate the transverse scaling exponent without taking into account the rescaling. This is an important result belonging to a frequently discussed issue in the analysis of fully-developed turbulence: the problem of possible differences in the scaling properties of longitudinal and transverse velocity increments in isotropic small scale turbulence. 

\subsection{Generalized K\'arm\'an equation}

\begin{figure}[tbp]
\begin{center}
\includegraphics[width=5in]{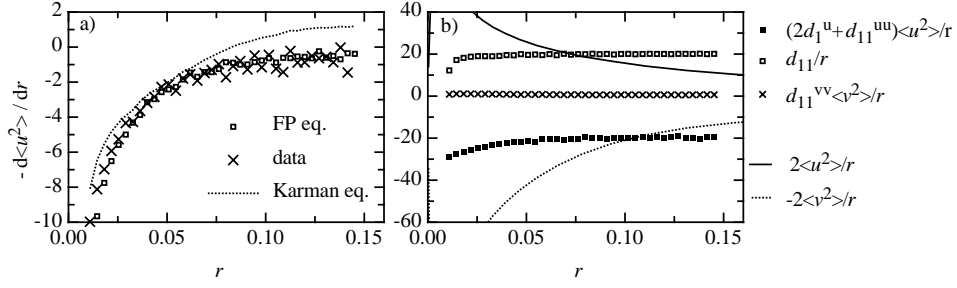}
\end{center}
\caption{a) Comparison of the left hand sides of the K\'arm\'an equation (\ref{Karman1}) with the structure function equation (\ref{FPKarman}) derived from the Fokker-Planck equation. Both equations are divided by $r$. Open squares: $-\partial_r \left< u_r^2\right>$ calculated with Eq. (\ref{FPKarman}); crosses: directly from data; dotted line: calculated from the K\'arm\'an equation. b) The different terms of (\ref{FPKarman}) (symbols) and (\ref{Karman1}) (lines) which contribute to a). The pre-factors of the structure functions of equation (\ref{FPKarman}) and (\ref{Karman1}) are different. \label{fig:MomentsEq}}
\end{figure}

The hierarchical equation (\ref{StrukturGl}) calculated from the Fokker-Planck equation is a kind of generalized K\'arm\'an equation: it is an equation which connects different moments. Let us compare the K\'arm\'an equation
\begin{eqnarray}\label{Karman1}
-r\frac{\partial \left< u_r^2\right>}{\partial r}=2\left< u_r^2\right>-2\left< v_r^2\right>
\end{eqnarray}
with the case $m=2,\;n=0$ of the hierarchy (\ref{StrukturGl}) 
\begin{equation}\label{FPKarman}
-r\frac{\partial \left< u_r^2\right>}{\partial r}=\left(2d_1^u+d_{11}^{uu}\right)\left< u_r^2\right> +d_{11}+d_{11}^{vv}\left< v_r^2\right>.
\end{equation}
Although the K\'arm\'an equation holds for Re$\to\infty$ and our result is obtained for quite moderate Reynolds number, the structure of both equations is remarkably similar, except for the additive term $d_{11}$. In Fig. \ref{fig:MomentsEq}a) is shown how well these equations reproduce $-\partial_r \left<u^2\right>$. The agreement of (\ref{FPKarman}) to the data is better than that of the K\'arm\'an equation because the flow seems to be slightly anisotropic. Thus we can also analyze anisotropic effects with the Fokker-Planck equation. We want to look in more detail at the differences by comparing the terms associated to the same structure functions on the right hand sides of the two equations, see Fig. \ref{fig:MomentsEq}b). It can be seen that the pre-factors of the structure functions are different for equation (\ref{Karman1}) and (\ref{FPKarman}): for example it is $\left(2d_1^u+d_{11}^{uu}\right)\left< u^2\right> \not\equiv -2\left< u^2\right>$. This reflects the different meaning of the two equations. Note that the anisotropy is clearer visible in Fig. \ref{fig:MomentsEq}a) than in Fig. \ref{fig:karman} because in \ref{fig:MomentsEq}a) the difference of two quantities of similar magnitude, right hand side of Eq. (\ref{Karman1}), is plotted and thereby the deviation to the isotropic case is more pronounced.

Whereas the K\'arm\'an equation connects only the second order structure functions, the Fokker-Planck equation is an equation also for higher orders. Furthermore its solution not only gives the relation between the structure functions but can even reproduce the structure functions itself. These informations are included in the $d$-coefficients. An additional difference is that the Fokker-Planck equation also includes the information of the multi-scale statistics.

We want to mention that also the second K\'arm\'an equation \cite{Monin75}
\begin{equation}\label{Karman2}
-r\frac{\partial \left< u_r^3\right>}{\partial r}=\left< u_r^3\right>-6\left< u_rv_r^2\right>
\end{equation}
has a similar structure as the hierarchical equation of the Fokker-Planck equation (Eq. (\ref{FPKarman}) for $m=3$ and $n=0$):
\begin{equation}
-r\frac{\partial \left< u_r^3 \right>}{\partial r}=\left(3d_1^u+3d_{11}^{uu}\right)\left< u_r^3\right> +3d_{11}^u\left< u_r^2\right> +3d_{11}^{vv}\left< u_rv_r^2 \right>.
\end{equation}

\subsection{Different `cascade speed'}

In Fig. \ref{fig:d_r} we have recognized by the drift and diffusion coefficients that the dominating differences between longitudinal and transverse increments can be found in a 3/2-times faster `cascade speed' of the transverse increments, which we have interpreted as a rescaling symmetry. This has some consequences for the increment statistics. First we want to examine how this symmetry becomes apparent in the structure functions. Then we discuss small differences between longitudinal and transverse increments, which exist beyond the differences in the different cascade speeds. At last we show that the rescaling is compatible to known results supporting our findings.

The rescaling-symmetry of the cascades speeds  can also be observed directly at the structure functions $\left<u_r^m\right>$ and $\left<v_r^m\right>$. This can be studied with aid of the hierarchical equation (\ref{StrukturGl}) and the coefficients in (\ref{ds}). We apply the rescaling to the transverse hierarchical equation and label all functions with a tilde, whose argument is multiplied by a factor 2/3. Then the equations for longitudinal and transverse structure functions read
\begin{eqnarray}
{-r\frac{\partial}{\partial r}\left< u_r^m\right>}&=&md_1^u\left< u_r^m\right>+\frac{m(m-1)}{2}d_{11}\left< u_r^{m-2}\right>+\frac{m(m-1)}{2}d_{11}^u\underline{\left< u_r^{m-1}\right>}\nonumber\\
&&+\frac{m(m-1)}{2}\underline{\underline{d_{11}^{uu}}}\left< u_r^{m}\right>+\frac{m(m-1)}{2}d_{11}^{vv}\left< u_r^{m-2}v_r^2\right>\label{StrukturGl1}\\
{-r\frac{\partial}{\partial r}{\widetilde{\left< v_r^m\right>}}}&=&m\widetilde{d}_2^v{\widetilde{\left< v_r^m\right>}}+\frac{m(m-1)}{2}\widetilde{d}_{22}{\widetilde{\left< v_r^{m-2}\right>}}+\frac{m(m-1)}{2}\widetilde{d}_{22}^u\underline{\widetilde{\left< u_r v_r^{m-2}\right>}}\nonumber\\
&&+\frac{m(m-1)}{2}\underline{\underline{\widetilde{d}_{22}^{vv}}}{\widetilde{\left< v_r^{m}\right>}}+\frac{m(m-1)}{2}\widetilde{d}_{22}^{uu}{\widetilde{\left< v_r^{m-2}u_r^2\right>}}.
\end{eqnarray}
Due to the 2/3-rescaling symmetry the coefficients $\tilde d_2^v$, $\tilde d_{22}$, $\tilde d_{11}^u$ and $\tilde d_{11}^{vv}$ can be replaced by the corresponding coefficients of the longitudinal equation; the only exception is the double underlined coefficient:
\begin{eqnarray}\label{StrukturGl2}
{-r\frac{\partial}{\partial r}{\widetilde{\left< v_r^m\right>}}}&=&md_1^u{\widetilde{\left< v_r^m\right>}}+\frac{m(m-1)}{2}d_{11}{\widetilde{\left< v_r^{m-2}\right>}}+\frac{m(m-1)}{2}d_{11}^u\underline{{\widetilde{\left< u_r v_r^{m-2}\right>}}}\nonumber\\
&&+\frac{m(m-1)}{2}\underline{\underline{\widetilde{d}_{22}^{vv}}}{\widetilde{\left< v_r^{m}\right>}}+\frac{m(m-1)}{2}d_{11}^{vv}{\widetilde{\left< v_r^{m-2}u_r^2\right>}}.
\end{eqnarray}
Equation (\ref{StrukturGl1}) and (\ref{StrukturGl2}) would have the same solution without the underlined terms as can be seen by comparison. In other words, without these terms the longitudinal and transverse structure functions also obey the 2/3-rescaling symmetry, $\left<|v^m(r)|\right>=\left<|u^m(3r/2)|\right>$, were we set $u_r\equiv u(r)$ and $v_r\equiv v(r)$.

To focus on the deviations of the 2/3-rescaling symmetry, we subtract (\ref{StrukturGl2}) from equation (\ref{StrukturGl1}), i.e. we examine the differences of $\left< u^m(r)\right>$ and $\left< v^m(2r/3)\right>\equiv\widetilde{\left< v^m(r)\right>}$:
\begin{eqnarray}\label{differences}
-\frac{\partial}{\partial \ln r}\left(\left< u_r^{m}\right>-\widetilde{\left< v_r^{m}\right>}\right)&=&md_1^u\left(\left< u_r^m\right>-\widetilde{\left< v_r^m\right>}\right)\\\nonumber&&+\frac{m(m-1)}{2}d_{11}\left({\left< u_r^{m-2}\right>-\widetilde{\left< v_r^{m-2}\right>}}\right)\\\nonumber&&+\frac{m(m-1)}{2}d_{11}^{vv}\left(\left< u_r^{m-2}v_r^2\right>-{\widetilde{\left< v_r^{m-2}u_r^2\right>}}\right)\\\nonumber&&+\frac{m(m-1)}{2}\underline{\left(d_{11}^{uu}\left< u_r^{m}\right>-\widetilde{d}_{22}^{vv}\widetilde{\left< v_r^{m}\right>}\right)}\\\nonumber&&+\frac{m(m-1)}{2}\underline{\underline{d_{11}^{u}\left(\left< u_r^{m-1}\right>-\widetilde{\left< u_rv_r^{m-2}\right>}\right)}}.
\end{eqnarray}
This equation is a differential equation for the differences $\left< u_r^{m}\right>-\widetilde{\left< v_r^{m}\right>}$ of the longitudinal and transverse structure functions. If the initial condition of this equation is zero, the solution would be zero without the underlined terms. Furthermore, because $md_1^u+m(m-1)d_{11}^{uu}/2<0$ for all order $m$ of interest, this equation has stable solutions and deviations in the initial conditions will decrease fast converging to the 2/3-rescaling apart from the discussed terms. The simply underlined term violates the 2/3-symmetry because of $d_{11}^{uu}(r)\neq d_{22}^{vv}(2r/3)$, the double underlined term violates it because of the not symmetric occurrence of the coefficient $d_{11}^u$ in the equation. For even $m$ the last term is the smallest one because of the odd moment and because of the small pre-factor, which is on the same magnitude as the quadratic term. The influence of the two last terms become larger with increasing order of the moments.

Although the deviations of the 2/3-rescaling are small, they have a crucial meaning for the small scale turbulence: without the terms $d_{11}^{uu}$ and $d_{22}^{vv}$ there were no intermittency, without the term $d_{11}^u$ there was no skewness of the longitudinal increments. To study how the form of the Kramers-Moyal coefficients leads to increasing intermittency with decreasing scales, we use the flatness $F_u\equiv\left< u_r^4\right>/\left< u_r^2\right>^2$. For a Gaussian distributed function, the flatness has the value 3 and deviations from this value can be interpreted as intermittency. If we differentiate the flatness with respect to $r$, we can relate the Fokker-Planck equation and the flatness (we shorten $S^{mn}:=\left< u_r^mv^n\right>$):
\begin{eqnarray}\label{intermittenz}
\lefteqn{r\frac{\partial F_u}{\partial r}=-4F_ud^{uu}_{11}+\frac{2d_{11}}{S^{20}}\left[F_u-3\right]+}\\&&\frac{2d^{vv}_{11}}{(S^{20})^{2}}\left[F_uS^{20}S^{02}-3S^{22}\right]-2\frac{1}{(S^{20})^2}\left[2S^{30}d^u_1+3S^{30}d^{u}_{11}\right]\nonumber\\
\lefteqn{r\frac{\partial F_v}{\partial r}=-4F_vd^{vv}_{22}+\frac{2d_{22}}{S^{02}}\left[F_v-3\right]+}\\&&\frac{2d^{uu}_{22}}{(S^{02})^{2}}\left[F_vS^{02}S^{20}-3S^{22}\right]-2\frac{1}{(S^{02})^2}\left[2S^{03}d^v_2+3S^{21}d^{u}_{22}\right].\nonumber\label{intermittenzb}
\end{eqnarray}
The similar structure of Eq. (\ref{intermittenz}) and (\ref{intermittenzb}) allows to discuss both equations together. The direction of the cascade is from large to smaller scales $r$, so that $\partial_r F_u<0$ means increasing intermittency towards smaller scales. The first term is the dominating negative term. The second term vanishes for large scales, because of the approximately Gaussian form of the distribution and is positive if intermittency increase. The third term simplifies to $\frac{12d_{11}^{vv}}{(S^{20})^{2}}\sigma_{12}^2$ because of the approximate Gaussian character for large scales with the covariance $\sigma_{12}^2$. This term describes the coupling between both components and is larger than zero, i.e. it acts against intermittency. The fourth term belongs to higher order corrections and describes the influence of the skewness onto the intermittency. To summarize, only the two quadratic terms $d_{11}^{uu}$ and $d_{22}^{vv}$ significantly produce intermittency of the longitudinal and the transverse component, respectively. 

Similar consideration can be done for the skewness $S_u:=\left< u_r^3\right>/(\left< u_r^2\right>)^{3/2}$, which can also be expressed by the Kramers-Moyal coefficients (it is $S_v\equiv 0$ because of reflection symmetry):
\begin{equation}\label{skewness}
r\frac{\partial S_u}{\partial r}=\frac{3}{2}\frac{S_u}{S^{20}}d_{11}-\frac{3}{(S^{20})^{1/2}}d_{11}^u-\frac{3}{2}S_ud_{11}^{uu}+3S_u\left(\frac{1}{2}\frac{S^{02}}{S^{20}}-\frac{S^{12}}{S^{30}}\right)d_{11}^{vv}.
\end{equation}
A positive derivative $\partial_r S_u>0$ means an increasing skewness for decreasing scales. The first term on the right hand side of Eq. (\ref{skewness}) is always negative. The second term is positive and is therefore necessary for a skewed distribution. The third term amplifies an existing skewness. Therefore the coefficient $d_{11}^u$ is essential for the skewness, but also the `intermittency' term $d_{11}^{uu}$ leads to an increasing skewness.

We can summarize the results in a short form:\\
{\em If the $r$-dependence of the transverse component is rescaled by a factor of $2/3$, then the only differences can be found in intermittency contributions and the skewness.}

\subsection{Compatibility to known results}

To support our findings of the rescaling symmetry, we show that it is compatible to the K\'arm\'an equation and to the ratio of Kolmogorov constants. These two aspects were already presented in \cite{siefert04}, but here we go in more detail and add some new aspects.

\begin{figure}[tbp]
\begin{center}
\includegraphics[width=5in]{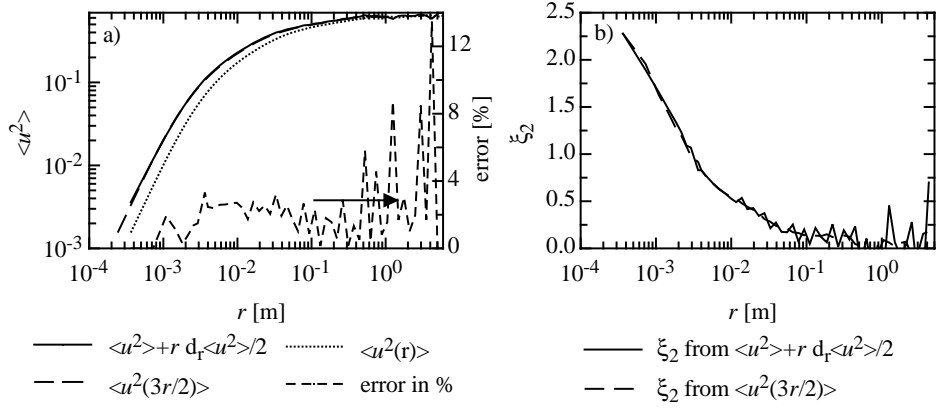}
\end{center}
\caption{a) The transverse second structure function calculated on the one hand by the K\'arm\'an equation (straight line) and on the other hand by using the 3/2-rescaling (broken line). Both curves deviate only within 4\% from another up to the integral scale. For comparison the longitudinal structure function is shown (dashed line). b) The local exponent for the transverse structure function calculated from the longitudinal structure function using the K\'arm\'an equation (straight line) and the 3/2-rescaling (broken line). The exponents are almost the same.
\label{fig:approxKarman} }
\end{figure}

The K\'arm\'an equation (\ref{karman1}) is consistent with the 2/3-rescaling symmetry, as one can see by interpreting the K\'arm\'an equation as a first-order Taylor expansion with the `small' quantity $r/2$:
\begin{eqnarray}\label{approxKarman}
\left< v^2(r)\right>&=&\left< u^2(r)\right>+\frac{1}{2}r\frac{\partial}{\partial r}\left< u^2(r)\right>\\&=&\left< u^2(r+\frac{1}{2}r)\right>-R=\left< u^2(\frac{3}{2}r)\right>-R.\nonumber
\end{eqnarray}
This equation correspond to the 2/3-rescaling symmetry of the second order structure functions except for the Lagrange remainder \cite{Bronstein} $R=r^2\,\partial_{\rho\rho}\left< u^2(\rho)\right>/8$ with $r<\rho<3r/2$. If we assume Kolmogorov-scaling \cite{kolmogorov41c} $\left<u^2\right>\propto r^{2/3}$ (intermittency effects for the second order structure function are negligible), the remainder $R=r^2\xi_2(\xi_2-1)\rho^{\xi_2-2}/8$ is below $2.1\%$ of $\left< u^2(3r/2)\right>$ for all $r$. An exact calculation yields that the relative error $R/\left< u^2(3r/2)\right>$ is \emph{independent} of $r$ and $R$ is $\approx 1.7\%$ of $\left< u^2(3r/2)\right>$, a value below the typical errors for $\left< u_r^2\right>$ of data from hot-wire anemometry. 
For the used data the approximation of $\left< u^2(r)\right>+r\,\partial_r\left< u^2(r)\right>/r$ by $\left< u^2(3r/2)\right>$ is valid within 4\% for all scales, see Fig. \ref{fig:approxKarman}a). Thus the 2/3-rescaling is a very good approximation of the K\'arm\'an equation. Because the only assumptions for these equation is a solenoidal and isotropic field,  the 2/3-rescaling is a property of isotropic turbulence, which should hold also for the limit Re$\to\infty$.

It is important to discuss the quality of the approximation (\ref{approxKarman}) with respect to scaling exponents.
The remainder $R=r^2/8\xi_2(\xi_2-1)\rho^{\xi_2-2}$ vanishes identically only for structure functions linear in $r$ ($\xi_2=1$). But as one can see from $\left<u^2(r)\right>+r/2\,\partial_r\left<u^2(r)\right>\propto r^{\xi_2}$ and $\left<u^2(3r/2)\right>\propto r^{\xi_2}$ the exponent is neither changed by the above approximation for pure scaling nor is the local exponent changed significantly for real data as one can see from figure \ref{fig:approxKarman}b). The reason is that the validity of the approximation depend only weak on the exponent. In reverse this does mean that the 2/3-rescaling does not distort the exponent strong.

The 2/3-rescaling can be related to the ratio of the Kolmogorov constants. Let us suppose that the structure functions scale with a power law, $\left< v^n(r)\right>=c_t^nr^{\xi^t_n}$ and $\left< u^n(r)\right>=c_l^nr^{\xi^l_n}$, even though our measured structure functions are still far away from showing an ideal scaling behavior \cite{renner02}. Note that the $c^n$ constants are related to the Kolmogorov constants but includes the energy dissipation $\left<\epsilon_r^{n/3}\right>$ and the enstrophy $\langle\Omega_r^{n/3}\rangle$. If we neglect the differences in intermittency and skewness, we can relate the structure functions according to the above mentioned rescaling: $\left< v^n(r)\right>=\left< u^n(3r/2)\right>=c_t^nr^{\xi^t_n}=c_l^n(3r/2)^{\xi^l_n}$. We end up with the relation $\xi^l_n=\xi^t_n$ and $c_t^n/c_l^n=\left(3/2\right)^{\xi^l_n}$. For $n=2$ and $n=4$ we obtain $c_t^2/c_l^2\approx 1.33$ and $c_t^4/c_l^4\approx 1.72$, which deviates less than  3\% from the value of $c_t^2/c_l^2 = 4/3$ and  $c_t^4/c_l^4=16/9$ given in \cite{antonia97c}. Thus  one needs only one constant 2/3 to explain the two ratios $c_t^2/c_l^2$ and $c_t^4/c_l^4$. We conclude that the 2/3-rescaling is the underlying relation between longitudinal and transverse Kolmogorov constants and therefore gives a forecast also for Kolmogorov-constants of higher orders.

\begin{figure}[tbp]
\begin{center}
\includegraphics[width=5in]{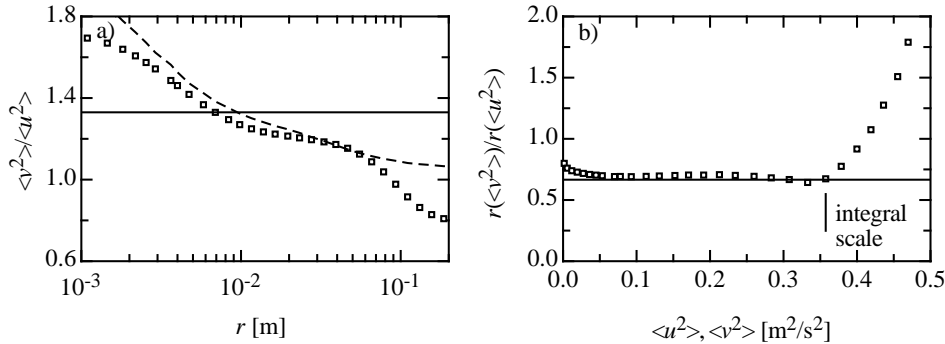}
\end{center}
\caption{a) The ratio $\left<v^2\right>/\left<u^2\right>$ (squares) calculated from data and the expected ratio if the structure functions would show Kolmogorov-scaling (straight line). The dashed line shows the isotropic case $\left<v^2\right>_\textrm{iso}/\left<u^2\right>$, where $\left<v^2\right>_\textrm{iso}$ is calculated with the K\'arm\'an equation. The just small differences between the two curves indicate that the deviations from the straight line can not be explained with anisotropy. b) The almost constant ratio $r(\left<v^2\right>=x)/r'(\left<u^2\right>=x)$ of the `inverse' of the structure functions close to the value 2/3 indicates that the differences between both structure functions can be found in their scale dependence (see text).
\label{fig:KolmogorovConst} }
\end{figure}

For finite Reynolds number the differences between the ratio of the Kolmogorov constants and the 2/3-rescaling become more obvious. The ratio $\left<v^2\right>/\left<u^2\right>=(c_t^nr^{\xi^t_n})/(c_l^nr^{\xi^l_n})=c_t/c_l=4/3$ is not fulfilled anymore, see Fig. \ref{fig:KolmogorovConst}a). Whereas for high Reynolds numbers this quantity can be interpreted as a ratio of amplitudes of the structure functions, this meaning gets lost for finite Reynold numbers. But if we treat the structure function as the independent variable and $r$ as the dependent variable, i.e. $r$ is a function of the structure functions, and calculate the ratio $r(\left<v^2\right>=x)/r'(\left<u^2\right>=x)$, we get an almost constant value close to 2/3, see Fig. \ref{fig:KolmogorovConst}b). This is a remarkably property: The constant ratio of the Kolmogorov constants are just a special case of the 2/3-rescaling. Whereas the former is valid only for high Reynolds number, the 2/3-rescaling is also a property of moderate Reynolds-number flows. A further consequence belongs to the discussion whether the Kolmogorov constants are universal or not \cite{barenblatt95,yeung97,praskovsky94}.  Even if the Kolmogorov constants are not universal, the ratio $r(\left<|v^m|\right>=x)/r'(\left<|u^m|\right>=x)$ seems to be universal if the turbulence is isotropic.

\subsection[Intermittency of longitudinal and transverse increments]{Intermittency and the coupling of longitudinal and transverse increments}

From the form of the Fokker-Planck equation it can be seen that intermittency is closely related to the interaction of longitudinal and transverse increments. More precisely, a necessary condition of intermittency is the mutual dependence of longitudinal and transversal increments. This can be deduced as follows.

First, we neglect the small differences in the intermittency and the skewness. If the 2/3-rescaling is applied, the form of the Fokker-Planck equation becomes symmetric with respect to exchange of $u_r$ and $v_r$. Moreover, because of the linear drift term and the rotational symmetric diffusion term, the Fokker-Planck equation becomes rotational symmetric and one can use circular coordinates with radius $\rho^2=u_r^2+v_r^2$. Thus rotational symmetric functions $F(\rho^2=u_r^2+v_r^2)$ solve the Fokker-Planck equation. 

To consider the question wether $u_r$ and $v_r$ are coupled, we construct a contradiction by assuming that $u_r$ and $v_r$ are independent, in the sense that $F(u_r^2,v_r^2)=F(u_r^2)F(v_r^2)$. This means for a solution $F$ the relation $F(\rho^2) = F(u_r^2+v_r^2)=F(u_r^2)F(v_r^2)$ holds. But this equation is only fulfilled for Gaussian distributed functions $F$ what is in contradiction to the fact that the turbulent flow is intermittent. Thus we conclude the two increments $u_r$ and $v_r$ depend of each other. Intermittency models has to take into account this coupling between the two increments and the one dimensional intermittency models has to be extended.

\subsection{Transverse scaling exponents}

In this section we focus on the wide debate about possible differences between the scaling exponents of longitudinal and transverse structure functions and consider the implications of the rescaling symmetry on it. In Fig. \ref{fig:ESSb} we have shown the recognized result that using ESS the transverse exponents are smaller than the longitudinal. Nevertheless, we use a new ansatz to reconsider this result and combine the differences of the longitudinal and transverse cascade speeds with ESS. 
 
Because the assumption for the longitudinal structure functions in ESS is $\langle| u^n(r)|\rangle\propto\left(r f_l(r)\right)^{\xi^l_n}$, see (\ref{ESSf}), the corresponding relation for the transverse increments has to be $\langle| v^n(r)|\rangle\propto\left(r f_t(r)\right)^{\xi^t_n}$ with a function $f_t$ different from $f_l$. Notice that the implicit assumption in ESS was $f_t\equiv f_l$, see (\ref{ESSa}) and (\ref{ESSb}). Using for the second order structure function the 3/2-rescaling found from the Fokker-Planck equation and assuming that the differences between skewness and intermittency are small for this order, we find a relation for the two functions: Starting with $\langle |v^2(r)|\rangle=\langle| u^2(3r/2)|\rangle$ we get $rf_t(r)=\left(3rf_l(3r/2)/2\right)^{\xi^l_2/\xi^t_2}$ (we have chosen the proportionality in ESS without loss of generality in such a way, that the equals sign holds). Because the intermittency corrections is small for the second order, it is $\xi^l_2\approx\xi^t_2$ and we get
\begin{equation}
	f_t(r)=3f_l(3r/2)/2.
\end{equation}
For arbitrary structure functions it then holds 
\begin{equation}\label{ESST}
\langle |v(r)|^n\rangle\propto\left(\frac{3}{2}rf_l(\frac{3}{2}r)\right)^{\xi^t_n}\propto\langle |u(\frac{3}{2}r)|^m\rangle^{\xi^{t}_{n}/\xi^{l}_{m}}=\langle| u(\frac{3}{2}r)|^3\rangle^{\xi^{t}_{n}}\;\;\textrm{(ESST)},
\end{equation}
which we call in the following ESST, extended self similarity for the transverse structure function.

\begin{figure}[tbp]
\begin{center}
\includegraphics[width=1.8in]{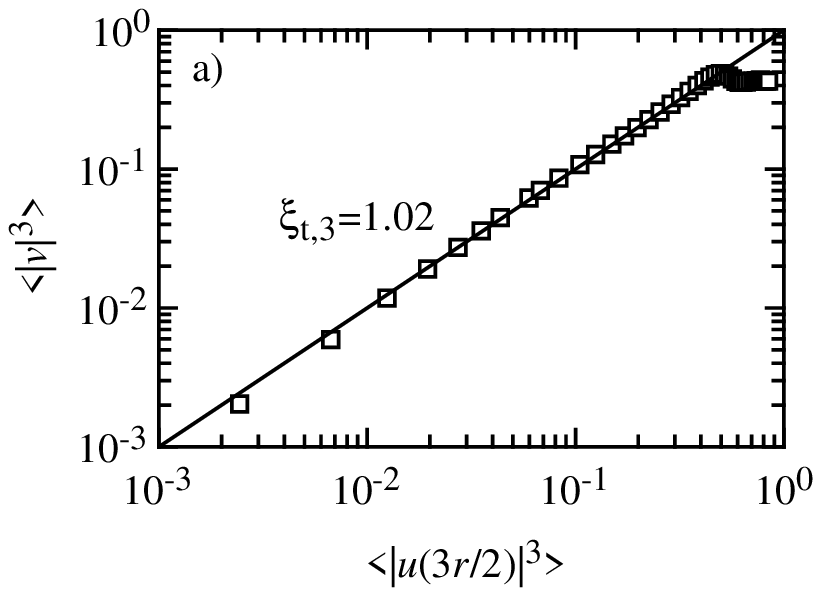}
\includegraphics[width=1.8in]{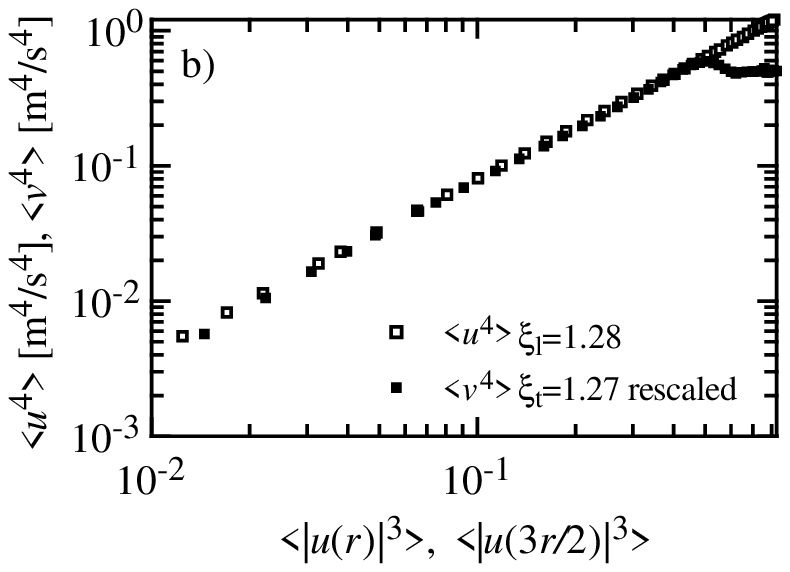}
\includegraphics[width=1.8in]{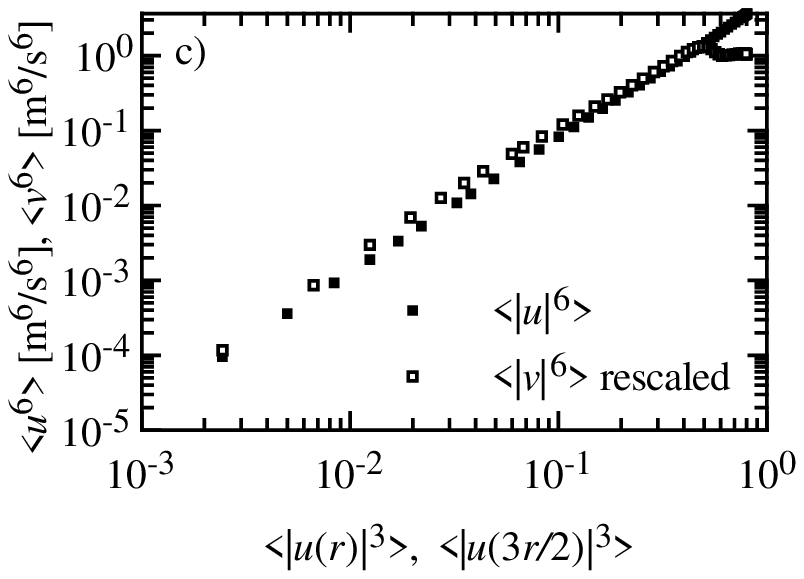}
\end{center}
\caption{The ESS representation of the longitudinal structure functions in comparison to the ESST representation of the transverse structure functions. a) The ESS representation of the transverse third order structure function. The exponent is $\xi^{t}_3=1.02$ and with that close to Kolmogorov's 4/5-law, see the line with the exponent 1. b) The fourth order structure functions and c) the sixth order structure functions. The transverse structure functions come significant closer to the longitudinal ones in comparison to Fig. \ref{fig:ESS}. \label{fig:ESST} }
\end{figure}

Fig. \ref{fig:ESST} shows the application of ESST to the transverse structure functions. As an remarkably result, the differences between both exponents vanish. Thus one scaling group is sufficient to characterize both increments. 

Notice that the differences between the exponents found with ESS are due to a none existing scaling behavior of the structure functions with $r$. It is evident that our rescaling does not change the exponents  in case of pure scaling behavior $\langle |v(r)|^n\rangle\propto (3r/2)^{\xi^t_n}\propto r^{\xi^t_n}$, which is expected to be valid if the Reynolds number goes to infinity (see also \cite{renner02}). 

\begin{figure}[tbp]
\begin{center}
\includegraphics[width=5in]{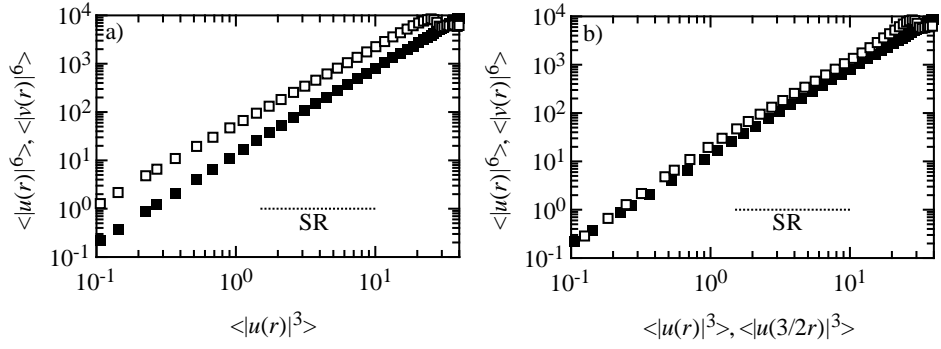}
\end{center}
\caption{The sixth order longitudinal (black squares) and transversal (withe squares) structure function in a) ESS representation and b) ESST representation.\label{fig:ESS6HRe}}
\end{figure}

\begin{figure}[tbp]
\begin{center}
\includegraphics[width=3.5in]{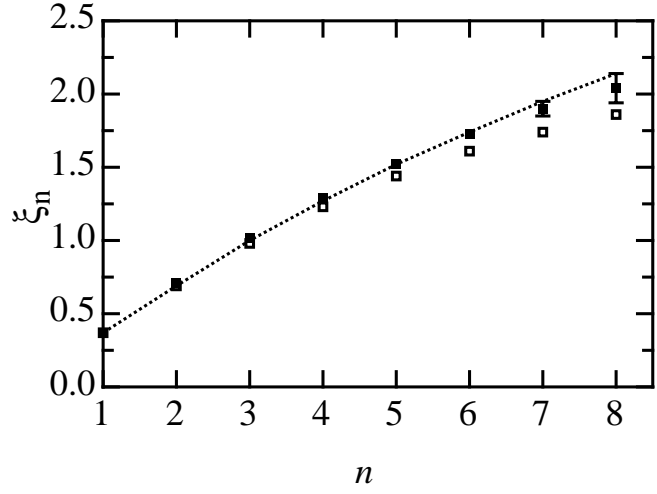}
\end{center}
\caption{Scaling exponents up to order 8 for longitudinal and transverse structure functions. Dashed line: longitudinal exponents. Open squares: transverse exponents estimated with ESS.  Filled squares: transverse exponents estimated with ESST. \label{fig:ExponentHRe}}
\end{figure}

Next we perform the analog analysis for our second data set with the higher Reynolds number $R_\lambda =550$. In Fig. \ref{fig:ESS6HRe}a) the sixth order structure function is plotted using ESS and in Fig.  \ref{fig:ESS6HRe}b) using ESST. Again we find that the differences in the scaling exponents vanish for ESST (ESS: $\xi^{l}_6=1.74\pm 0.03$, $\xi^{t}_6=1.60\pm 0.03$; ESST: $\xi^{l}_6=1.74\pm 0.03$, $\xi^{t}_6=1.75\pm 0.03$). In Fig. \ref{fig:ExponentHRe} the exponent $\xi^l_n$ as well as $\xi^t_n$ is plotted up to order 8. The differences between the exponents ESS vanish  within the uncertainties applying ESST instead of ESS.

\begin{figure}[tbp]
\begin{center}
\includegraphics[width=5in]{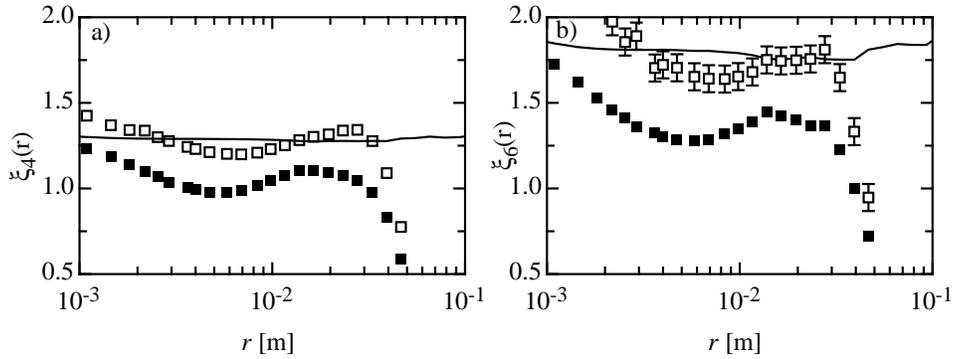}
\end{center}
\caption{The local exponent $\xi^{\alpha}_n=\partial \log\langle \alpha^n\rangle/\partial \log\langle|\beta|^3\rangle$ with $\alpha=u(r)$ and $\beta=u(r)$ (straight line); $\alpha=v(r)$ and $\beta=u(r)$ (black squares); $\alpha=v(r)$ and $\beta=u(3r/2)$ (squares). a) Local exponent of the fourth order structure function and b) for the sixth order structure function. \label{fig:localExponent} }
\end{figure}

To see in more detail the influence of ESST to the exponents, in Fig. \ref{fig:localExponent} the local exponents $\xi^{\alpha}_n(r)=\partial \log\langle \alpha^n\rangle/\partial \log\langle|u|^3\rangle$ ($\alpha=u,v\;\;n=4,6$) are shown. The use of ESS for the longitudinal exponents results in an almost constant local exponent. The transverse exponents have oscillations, which were explained by log-periodic oscillations. If ESST is applied, the differences between longitudinal and transverse exponents vanish within these oscillations.

\section{Conclusion}

We have analyzed the differences and similarities between longitudinal and transverse increments. Differences in their structure functions and probability distributions were found, which were observed already from other groups. To look in more detail at these differences we have used a method to estimate by pure data analysis a phenomenological Fokker-Planck equation and have extended this analysis to both, longitudinal and transverse increments. We have checked carefully the mathematical prerequisites of this method and found that they are well fulfilled for our data. Then we have estimated the drift and diffusion coefficient of the Fokker-Planck equation directly from the data. The solution of the Fokker-Planck equation with these coefficients reproduces well the probability distributions and the structure functions. Thus, the statistics of the joint probability of longitudinal and transverse increments is encoded in the drift and diffusion coefficients. These new quantities for analyzing the turbulent flow determine a stochastic process in $r$, which can be interpreted as a `cascading process'. The statistics of both increments are decoupled in the deterministic drift vector but are coupled via the nonlinear and no diagonal diffusion matrix. 

A remarkably result is that in a first approximation the longitudinal and transverse drift and diffusion coefficients can be transformed to each other by a simple rescaling, namely, by multiplying the scale of the longitudinal  increments with the factor 3/2. With this rescaling the frequently discussed issue of the differences between longitudinal and transverse structure functions can be explained. In this context the extended self similarity method (ESS) has to be changed for the transverse structure functions according to the rescaling symmetry, leading to an extended self-similarity for the \emph{transverse} increments (ESST). From it, we get the remarkably result that the previous found differences in the scaling exponents vanish. Accordingly, the seemingly differences between the exponents were a result of the non-scaling behavior. Thus it is evident that with increasing Reynolds number the differences in the scaling exponents diminish, as more and more ideal scaling behavior occurs. It is interesting to note that the 3/2 rescaling, which we interpret as different speed of the longitudinal and transverse cascade, is compatible with the K\'arm\'an equation. This leads to the proposal that the 3/2 rescaling is valid for large Reynolds number Re$\to \infty$.
 
 Besides the features of structure functions our method of analyzing the statistics by means of a Fokker-Planck equation, provides further insights in the complexity of the cascading process. Access to stochastic properties on multi-scales and the interaction between both increments are  given. In this closer look principal differences between longtidudinal and transverse increments can be identified, which are not grasped by the investigation with structure functions. The physical meaning of these differences has to be clarified for an extended understanding of the small scale turbulence.

We acknowledge fruitful discussions with R. Friedrich, A. Naert and teamwork with S. L\"uck and F. Durst. This work was supported by the DFG-grant Pe 478/9.
 
\section{Appendix A: Extended self similarity}\label{appendixa}

Benzi \emph{et. al.} \cite{benzi93b,benzi95} have noticed that structure functions show extended self similarity (ESS) and that this property can be used to estimate scaling exponents for moderate Reynolds number for which a scaling regime is not well-developed. The basic assumption of ESS is that the structure functions of different order have a similar shape, which can be written as 
\begin{equation}
\left<|u_r|^n\right>\propto(rf(r))^{\xi_n^l}\label{ESSf}
\end{equation}
 with the common function $f(r)$. Additionaly it was shown that $\left<|u^3|\right>$ and $\left<u^3\right>$ have the same scaling properties and if one choses according to Kolmogorov's 4/5-law $\xi_3^l=1$ one can write ESS in the form 
\begin{equation}
\left<|u_r|^n\right>\propto\left<|u_r|^3\right>^{\xi_n^l}\label{ESSa}.
\end{equation}
Using this expression, the scaling range extends down to $3\eta$--$5\eta$ and the exponent is universal, i.e. does not depend on the Reynolds number.

For the transverse structure functions ESS was applied in two different ways. Some groups have plotted $\langle v_r^n\rangle\propto \langle|v_r|^3\rangle^{\xi_n^t}$ \cite{camussi96b,noullez97}. But in recent time it has been argued that 
\begin{equation}
\langle |v_r|^n\rangle\propto \langle|u_r|^3\rangle^{\xi_n^t}\label{ESSb}
\end{equation}
 is theoretical more justified because Kolmogorov's 4/5-law gives an exact prediction for the longitudinal third order structure function and therefore $\langle|u_r|^3\rangle$ is a good point of reference \cite{pearson01,dhruva97,antonia97a,zhou00,chen97a}.

\section{Appendix B: Left bounded increments}\label{appendix}

\begin{figure}[tbp]
\begin{center}
\includegraphics[width=5in]{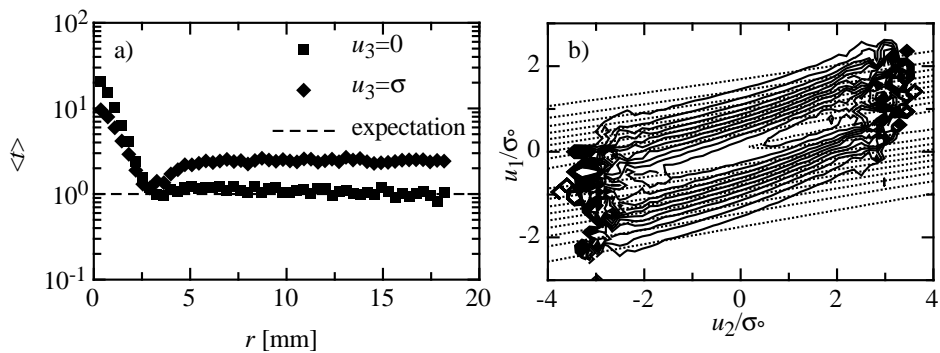}
\end{center}
\caption{Properties of the left-bounded increments. a) The same plot as in Fig. \ref{fig:markov3} but for left-bounded increments. For the case $u_3=\sigma$ the Markov properties are violated. b) The solution of the Fokker-Planck equation (dotted line) in comparison to the conditional  probability distribution (full line) estimated from data. The conditioned pdfs can not be described by this Fokker-Planck equation.\label{fig:markov5}}
\end{figure}

\begin{figure}[tbp]
\begin{center}
\includegraphics[width=5in]{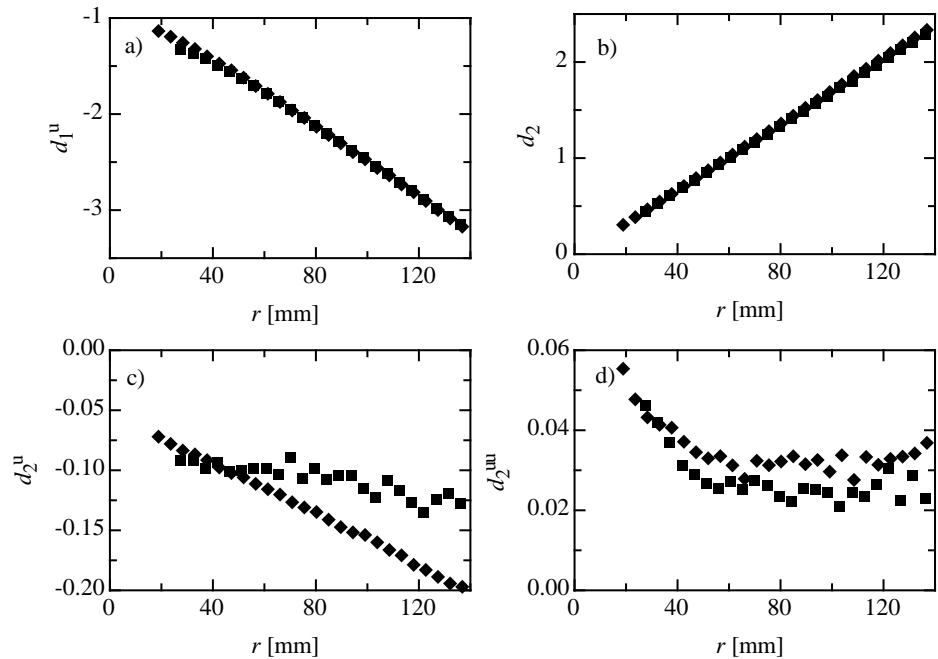}
\end{center}
\caption{The expansion coefficients of the Kramers-Moyal coefficients according to eq. (\ref{ds1}) for centered and left-bounded increments. The coefficients for the centered increments (black squares) and the left-bounded increments (white squares) are equal for the smallest order in $u$ and $v$, respectively, see a) and b). For the linear and quadratic term of the diffusion coefficients the deviations are significant, see c) and d)). But because these two coefficients have only a small contribution to the drift and diffusion coefficients, the effective influence on the Fokker-Planck equation is nevertheless weak.
\label{fig:CoeffLeftCentered}}
\end{figure}

In the case of multi-point statistics with $n$ increments on different scales $r_1,\,\dots,\,r_n$, one has to define the relative position of increments. A general definition is $u_r(\alpha)={\bf e}\cdot\left[ {\bf U}({\bf x}+\alpha{\bf r})-{\bf U}({\bf x}-(1-\alpha){\bf r})\right]$ with $\alpha \in[0,1]$. We call the case $\alpha=1$ left-bounded increments and the case $\alpha=1/2$ centered increments. The left-bounded increments are the usual way to define the relative position of several increments, see for example \cite{kolmogorov41c}. Although, for this increment definition, the Markov properties are violated as one can see in Fig. \ref{fig:markov5}a). For $u_3=0$ the Markov properties are fulfilled. For $u_3=\sigma$ we get Markov properties around the expected Markov-length, but for larger scales the Markov properties are violated. Nevertheless, the violation of the Markov properties does not affect the Kramers-Moyal coefficients very much; even the deviations in $d_2^u$ does not effect the total Fokker-Planck equation very much, because this term is small in comparison to the others, see Fig. \ref{fig:CoeffLeftCentered}. The violated Markov properties become more obviously, if the conditioned moments estimated from the Fokker-Planck equation are compared with those from the data, see Fig. \ref{fig:markov5}b). It can be seen that the left-bounded increments does not describe the conditional moments. Therefore we use the centered increments for our analysis. For a discussion on the background of random numbers and random walks see Ref. \cite{waechter04}


\bibliographystyle{plain}
\bibliography{lit}

\end{document}